\documentclass[reprint,amsmath,amssymb, aps, prx,superscriptaddress, longbibliography]{revtex4-2}
\usepackage{graphicx}
\usepackage{amsmath}
\usepackage{braket}
\usepackage{tikz}
\usepackage{float}
\usetikzlibrary{positioning}
\usepackage{soul,xcolor}
\usepackage[export]{adjustbox}
\usepackage{placeins}
\usepackage[scr=boondox]{mathalpha}
\usepackage{ulem}
\usepackage{ragged2e}
\usepackage{amsthm}

\usepackage[linktocpage]{hyperref}
\hypersetup{colorlinks=true,citecolor=blue,linkcolor=blue, urlcolor=blue, breaklinks=true}

\newcommand{\Rom}[1]{\uppercase\expandafter{\romannumeral #1\relax}}

\newtheorem*{assumption*}{\assumptionnumber}
\providecommand{\assumptionnumber}{}


\theoremstyle{definition}

\theoremstyle{theorem}

\theoremstyle{corollary}

\theoremstyle{lemma}

\theoremstyle{Proposition}

\theoremstyle{definition}

\makeatletter

\newcommand{\Rmnum}[1]{\expandafter\@slowromancap\romannumeral #1@}
\makeatother
\newcommand{\imth}{\hspace{1pt}\mathrm{i}\hspace{1pt}}

\newcommand{\bea}{\begin{eqnarray}}
\newcommand{\eea}{\end{eqnarray}}
\newcommand{\bct}{\begin{center}}
\newcommand{\ect}{\end{center}}
\newcommand{\bpm}{\begin{pmatrix}}
\newcommand{\epm}{\end{pmatrix}}
\newcommand{\bal}{\begin{aligned}}
\newcommand{\eal}{\end{aligned}}

\newcommand{\expval}[1]{\langle{#1}\rangle}

\newcommand{\dket}[1]{|{#1}\rangle}

\begin{document}
	\title{Optical and Raman selection rules for odd-parity clean superconductors}
 \author{Shuangyuan Lu}
 \author{Xu Yang}
	\author{Yuan-Ming Lu}
	\affiliation{Department of Physics, The Ohio State University, Columbus, OH 43210, USA}
	
	\date{\today}
	
	\begin{abstract}
        We derive selection rules in optical absorption and Raman scattering spectra, that can determine the parity of pairing order parameters under inversion symmetry in two classes of \emph{clean} superconductors: (i) chiral superconductors with strong spin-orbit couplings, (ii) singlet superconductors with negligible spin-orbit couplings. Experimentally, the inversion parity of pair wave functions can be determined by comparing the ``optical gap'' $\Delta_\text{op}$ in Raman and optical spectroscopy and the ``thermodynamic gap'' $2\Delta$ in specific heat measurements, and the selection rules apply when $\Delta_\text{op}>2\Delta$. We demonstrate the selection rules in superconductivity in models of (i) doped Weyl semimetals and (ii) doped graphene. Our derivation is based on the relation between pairing symmetry and fermion projective symmetry group of a superconductor. We further derive similar selection rules for two-dimensional superconductors with 2-fold rotational symmetry, and discuss how they apply to the superconducting state in magic-angle twisted bilayer graphene. 
    \end{abstract}

	\maketitle

\section{{Introduction}}
As a best known quantum phenomenon in the macroscopic scale, superconductivity has many important technological applications ranging from quantum sensing\cite{Degen2017} and quantum computing\cite{Kjaergaard2023} to improving energy efficiency\cite{MacManus-Driscoll2021}. Spontaneously breaking the $U(1)$ charge conservation symmetry in the material, superconductors (SCs) exhibit a Landau-type off-diagonal long range order, characterized by a pairing order parameter $\Delta(\mathbf{k})$ in the momentum space (labeled by crystal momentum $\mathbf{k}$), also known as the pair wave function\cite{Annett1990,Sigrist1991,Tsuei2000}. Most metals and alloys are conventional superconductors, where superconductivity is induced by the electron-phonon coupling and well described by the BCS theory, with an isotropic $s$-wave spin-singlet pairing order parameter. On the other hand, other types of interactions can give rise to non-$s$-wave {unconventional superconductivity}\cite{Sigrist1991} with many desirable properties, such as high temperature $d$-wave superconductivity in cuprates with strong Coulomb repulsions\cite{Damascelli2003,Lee2006,Taillefer2010,Damascelli2003,Armitage2010}, and topological superconductivity in materials with strong spin-orbit interactions\cite{Ando2015,Sato2017}. Unique applications can arise from unconventional superconductivity, such as fault-tolerant topological qubits based on Majorana zero modes\cite{Nayak2008,Alicea2012,Beenakker2013,Lutchyn2018} (MZMs).

However, to identify an unconventional superconductor, the experimental determination of the pairing symmetry has been a challenging task for almost every candidate material\cite{Annett1990,Sigrist1991,VanHarlingen1995,Tsuei2000}, for the following reason. To determine the symmetry of the complex pair wave function $\Delta(\mathbf{k})$, one needs information of both its magnitude $|\Delta(\mathbf{k})|$ and the phase $\arg\big[\Delta(\mathbf{k})\big]$. Most common experimental probes are believed to be only sensitive to the magnitude of the order parameter, such as penetration depth, specific heat, thermal transport, angle-resolved photoemission and NMR spectroscopy\cite{Scalapino1995,VanHarlingen1995,Tsuei2000}. On the other hand, phase-sensitive measurements, able to probe the relative phase of the order parameter as a function of $\vec k$-space direction, require more complicated devices and measurements, such as SQUID interferometry and tricrystal/tetracrystal magnetometry\cite{VanHarlingen1995,Tsuei2000}, which are not easily accessible to many candidate materials. Therefore, new experimental probes that can sharply determine the symmetry representation of the pairing order parameter $\Delta(\mathbf{k})$ are highly desirable.

As one simplest manifestation of pairing symmetry, in the presence of inversion symmetry, the pairing order parameter can either be an even or an odd function of $\vec k$, corresponding to (conventional) even-parity and (unconventional) odd-parity superconductivity. Odd-parity superconductors are often associated with exotic physical properties, such as MZMs in the vortex core of chiral $p$-wave superconductors in two dimensions (2d)\cite{Read2000}. In this work, we derive a set of selection rules in optical absorption and in Raman spectroscopy, for the particle-hole continuum of Bogoliubov quasiparticle excitations in a superconductor, that allows us to sharply distinguish odd-parity from even-parity superconductors. These selection rules are derived using the correspondence between the pairing symmetry and fermion Projective Symmetry Group (PSG) in a SC, recently established in Ref.\cite{Yang2024}. We demonstrate how to use the selection rules to detect the parity of pair wave functions in two systems: (i) chiral superconductors in doped Weyl semimetals, as an example of chiral superconductors with strong spin-orbit couplings (SOCs), (ii) singlet superconductors in doped graphene, as an example of singlet superconductors with $SU(2)$ spin rotational symmetries. For superconductors in 2d materials, we establish similar selection rules for the parity of pair  wavefunction under a 2-fold rotational symmetry $C_{2,z}$, and discuss how they apply to superconductors in magic-angle twisted bilayer graphenes (MATBG)\cite{Cao2018}.

{\emph{Symmetry actions on Bogoliubov quasiparticles}}~~~In the optical absorption and Raman spectroscopies of a SC, the dominating electronic contribution comes from a pair of Bogoliubov quasiparticles with opposite momenta, known as the ``particle-hole continuum'' of a SC. To obtain their selection rules under a crystalline symmetry such as inversion, we need to understand how the Bogoliubov quasiparticles transform under crystalline symmetry operations. This is captured by the following Bogoliubov de Gennes (BdG) Hamiltonian for the SC phase:
\bea\label{BdG ham}
&\hat H_\text{BdG}=\hat H_0+(\hat H_\text{pair}+~h.c.),\\
&\hat H_0=\sum_{\alpha\, \beta ;\mathbf{k}} \hat{c}^\dagger_{\mathbf{k}\alpha} h_{\alpha \beta}(\mathbf{k}) \hat{c}_{\mathbf{k}\beta}\\
&\hat{H}_{\text{pair}} = \sum_{\alpha\, \beta; \mathbf{k}} \hat{c}^\dagger_{\mathbf{k}\alpha} \Delta_{\alpha \beta}(\mathbf{k}) \hat{c}^\dagger_{-\mathbf{k}\beta}
\eea
where $\hat H_0$ describes the normal-state band structure $h_{\alpha,\beta}({\mathbf{k}})$ of electrons, and $\hat H_\text{pair}$ describes the Cooper pairing of electrons. We use $\alpha,\beta$ to generally denote the spin, orbital and sublattice indices of electrons. Under a crystalline symmetry operation $\hat g$, the electron transforms as $\hat{g}\, c_{\mathbf{k}\, \alpha} \hat{g}^{-1}\; = \; [ U^g_0(\mathbf{k}) ]^\dagger _{\alpha \beta} \; \hat{c}_{g\mathbf{k}\, \beta}$ where $U^g_0(\mathbf{k})$ is a unitary matrix. Although the normal-state band structure preserves the $\hat g$ symmetry as $[\hat g,\hat H_0]=0$, the pairing term $\hat H_\text{pair}$ in an unconventional SC is generally NOT invariant under crystal symmetry $\hat g$\cite{Yang2024,Shiozaki2022,Ahn2021}.

In this paper we restrict ourselves to the SCs without spontaneous breaking of spin rotational symmetries, i.e. the normal state and SC shares the same global (spin rotation) symmetry group. In this case, due to the broken $U(1)$ charge symmetry, the pairing term $\hat H_\text{pair}$ can acquire a phase $e^{\imth\Phi_g}$ under crystal symmetry $\hat g$\cite{Ahn2021,Shiozaki2022,Yang2024}:
\bea
U^g_0(\mathbf{k}) \Delta(\mathbf{k}) \left[ U_0^g(-\mathbf{k})\right]^T    \,=\, e^{i\Phi_g}\, \Delta(\hat g\mathbf{k})
\eea
The phase factors $\{e^{\imth\Phi_g}|\hat g\in X\}$ form a one-dimensional irreducible representation (irrep) of the crystal symmetry group $X$, satisfying $e^{\imth(\Phi_g+\Phi_h)}=e^{\imth\Phi_{gh}}$ for any $g,h\in X$. 

For the order-2 inversion symmetry $\hat \Gamma$ of interests to this work, we have $e^{\imth\Phi_\Gamma}=\pm1$ and hence
\bea
U^\Gamma_0(\mathbf{k}) \Delta(\mathbf{k}) \left[ U_0^\Gamma(-\mathbf{k})\right]^T    \,=\pm\Delta(-\mathbf{k}).
\eea
where $\pm$ signs correspond to even- and odd-parity under inversion symmetry respectively. Note that due to nontrivial transformations of the pairing wavefunction $\Delta(\mathbf{k})$ shown above, the BdG Hamiltonian is not invariant under the normal-state symmetry $\hat g$ anymore. Instead, $\hat H_\text{BdG}$ preserves a combination $\hat g^\prime$ of normal state crystal symmetry $\hat g\in X$ and a $U(1)$ charge rotation $e^{-\imth\Phi_g\hat F/2}$ where $\hat F$ is the total fermion number:
\bea\notag
&\hat g^\prime \hat H_\text{BdG}\big(\hat g^{\prime}\big)^{-1}=\hat H_\text{BdG},~~~\hat g^\prime=e^{-\imth\Phi_g\hat F/2}\hat g;\\
&\hat{g}^\prime\, c_{\mathbf{k},\alpha}\big(\hat{g}^\prime\big)^{-1}=e^{-\imth\Phi_g/2}[ U^g_0(\mathbf{k}) ]^\dagger _{\alpha \beta}\hat{c}_{\hat g\mathbf{k},\beta}.
\eea 
In the case of order-2 inversion symmetry $\Gamma$, we have 
\bea\label{psg:inversion}
\big(\Gamma^\prime\big)^2=e^{-\imth\Phi_\Gamma\hat F}=(\pm1)^{\hat F}
\eea
In other words, inversion squares to $\pm1$ when acting on a fermion operator in a SC with an even/odd-parity under inversion symmetry. They correspond to two different fermion PSGs\cite{Yang2024} with distinct physical properties. In particular, we consider Bogoliubov quasiparticles (BQPs) $\{\gamma_{\mathbf{k},a}\}$ of the BdG Hamiltonian (\ref{BdG ham}): 
\bea
\hat H_\text{BdG}=\sum_{\mathbf{k},a}E_a(\mathbf{k})\gamma^\dagger_{\mathbf{k},a}\gamma_{\mathbf{k},a},~~~E_a({\bf k})\geq0.
\eea
where $a$ generally labels the spin/band indices of BQPs. 
 
In optical absorption and Raman spectroscopy experiments, the dominant electronic contribution comes from a pair of BQPs with opposite momenta. Considering inversion symmetry $\Gamma^\prime$ in (\ref{psg:inversion}) of the SC phase, we can always choose a gauge so that 
\bea
\Gamma^\prime\bpm\gamma^\dagger_{\mathbf{k},a}\\ \gamma^\dagger_{-\mathbf{k},a}\epm(\Gamma^\prime)^{-1}=\bpm\gamma^\dagger_{-\mathbf{k},a}\\ \pm\gamma^\dagger_{\mathbf{k},a}\epm
\eea
This relation and fermi statistics together lead to a gauge-invariant inversion eigenvalue for a pair of BQPs with opposite momenta:
\bea\label{BQP pair:chiral}
\Gamma^\prime(\gamma^\dagger_{\mathbf{k},a}\gamma^\dagger_{-\mathbf{k},a})(\Gamma^\prime)^{-1}=\mp(\gamma^\dagger_{\mathbf{k},a}\gamma^\dagger_{-\mathbf{k},a})
\eea
In a chiral SC with strong spin-orbit couplings (SOCs), there is no Kramers degeneracy for BQPs at a generic momentum, and therefore the BQP pair with the lowest energy comes from the same Bogoliubov band $a=0$. Therefore inversion quantum number (\ref{BQP pair:chiral}) directly applies to the low frequency spectroscopy of a chiral SC. On the other hand, in a singlet SC with $SU(2)$ spin rotational symmetries, each Bogoliubov band has a 2-fold spin degeneracy $\alpha,\beta=\uparrow,\downarrow$ at every momentum. In this case, the spin-singlet BQP pair with the lowest energy has the following gauge-invariant inversion eigenvalue:
\bea\label{BQP pair:singlet}
\Gamma^\prime(\epsilon^{\alpha\beta}\gamma^\dagger_{\mathbf{k},0,\alpha}\gamma^\dagger_{-\mathbf{k},0,\beta})(\Gamma^\prime)^{-1}=\pm(\epsilon^{\alpha\beta}\gamma^\dagger_{\mathbf{k},0,\alpha}\gamma^\dagger_{-\mathbf{k},0,\beta})~~
\eea
Here, band index $0$ in the subscript means the lowest BdG band with a non-negative energy. Below we show how the gauge-invariant inversion eigenvalues in (\ref{BQP pair:chiral})-(\ref{BQP pair:singlet}) lead to selection rules in optical absorption and Raman spectroscopy, which can be used to detect the inversion parity of pairing order parameters. 

\section{{Optical absorption spectroscopy}}

It was believed that optical spectroscopy via inelastic light scattering is sensitive only to the magnitude of the pairing order parameter\cite{Tsuei2000,Devereaux2007}. Although Higgs modes have been proposed to characterize and differentiate different pairing symmetries in unconventional superconductors\cite{Schwarz2020}, they have proven difficult to be observed in superconductors due to the nonlinear
light-Higgs coupling\cite{Shimano2023}. On the other hand, the particle-hole continuum above the $2\Delta$ threshold, created by breaking a Cooper pair and exciting two Bogoliubov quasiparticles, is the most important electronic responses to inelastic optical probes in a superconductor\cite{Devereaux2007}. Below we demonstrate that the aforementioned symmetry transformations on BQPs lead to distinct optical selection rules for the particle-hole continuum in SCs. 

The dominant component of light-matter interaction is the coupling of vector potential $\mathbf{A}$ of light with the particle current operator $\hat{\mathbf{j}}$ of the system:
\bea \label{eq.H_int}
\hat H_\text{int}=-e\mathbf{A}(t)\cdot\hat{\mathbf{j}}
\eea
The optical absorption rate is given by the transition rate in time-dependent perturbation theory, where the 1st-order perturbation theory leads to
\bea \label{eq.absorption}\notag
&\sum_fw_{i\rightarrow f}=\frac{2\pi}\hbar\sum_f|\expval{f|\hat H_\text{int}|i}|^2\delta(E_f-E_i-\omega)\\
\label{eq.absorption}&= \frac{\pi e^2}{\epsilon_0 \omega}\sum_f|\expval{i|\mathbf{e}\cdot\hat{\mathbf{j}}|f}|^2\delta(\omega+E_i-E_f)\\
\label{eq.absorption.response}&=\frac{\hbar}{\epsilon_0}\omega\cdot\text{Im}\chi(\omega)=\frac\hbar{\epsilon_0}\text{Re}\sigma(\omega)
\eea
where $\mathbf{e}$ is the polarization vector of the light, $\chi(\omega)=\chi^{\prime}+\imth\chi^{\prime\prime}$ and $\sigma(\omega)$ are the electric susceptibility and optical conductivity in linear response theory. At zero temperature, the initial state $\dket{i}$ is the SC ground state $\dket{0}$, and the final state $\dket{f}$ is obtained by creating a zero-momentum BQP pair on the ground state so that 
\begin{align}\notag
   &\frac{\hbar}{\epsilon_0}\omega~\text{Im}\chi(T=0,\omega)= \frac{\pi e^2}{2\epsilon_0\omega}\sum_{\mathbf{k},a,b}|\expval{0|\mathbf{e}\cdot\hat{\mathbf{j}} ~(\gamma^\dagger_{\mathbf{k},a}\gamma^\dagger_{-\mathbf{k},b})|0}|^2\\
   &\cdot\delta\big(\omega-E_a(\mathbf{k})-E_b(-\mathbf{k})\big)+\cdots
\end{align}
where $\cdots$ stands for other contributions from e.g. Higgs modes or 4 Bogoliubov quasiparticles. If we label the bottom of the lowest and 2nd lowest BQP bands as $\Delta$ and  $\Delta^\prime$ respectively with $\Delta<\Delta^\prime$, for the following frequency range
\bea\label{freq. range}
\omega<\Delta+\Delta^\prime
\eea
only a pair of BQPs from the lowest Bogoliubov band $a=b=0$ contributes, which corresponds to 
\bea
\dket{f}=\gamma^\dagger_{\mathbf{k},0}\gamma^\dagger_{-\mathbf{k},0}\dket{0}
\eea
in a chiral SC with a strong SOC, or
\bea
\dket{f}=\epsilon^{\alpha\beta}\gamma^\dagger_{\mathbf{k},0,\alpha}\gamma^\dagger_{-\mathbf{k},0,\beta}\dket{0}
\eea
in a singlet SC with $SU(2)$ symmetry. Since the current operator $\hat{\mathbf{j}}$ is odd under inversion symmetry
\bea
\Gamma^\prime\hat{\mathbf{j}}(\Gamma^\prime)^{-1}=-\hat{\mathbf{j}}
\eea
Eqs. (\ref{BQP pair:chiral})-(\ref{BQP pair:singlet}) immediately lead to optical absorption selection rules summarized in Table \ref{tab:chiral SC w/ SOC} and \ref{tab:singlet SC w/o SOC}, for chiral SCs with SOC and singlet SCs without SOC respectively. It is worth mentioning that the above absoprtion selection rules recover one well-known conclusion in the single-band ($\Delta^\prime\rightarrow\infty$) limit: there is no optical absorption for $\omega>2\Delta$ in single-band clean $s$-wave SCs\cite{Mahan2000B}. 

\onecolumngrid

\begin{table}[h]
    \centering
    \begin{tabular}{|c|c|c|c|c|}
    \hline
         Parity of $\Delta(\mathbf{k})$ & $(\hat \Gamma^\prime)^2$ & Parity of $\gamma^\dagger_{\mathbf{k},a}\gamma^\dagger_{-\mathbf{k},a}$ & Im$\chi(2\Delta<\omega<\Delta+\Delta^\prime)$&$I_\text{Raman}(2\Delta<\omega<\Delta+\Delta^\prime)$\\
    \hline
        Even&+1& Odd&Nonzero&0\\
        \hline
        Odd&-1&Even&0&Nonzero\\
    \hline
    \end{tabular}
    \caption{Selection rules for the parity of pair wavefunctions under inversion symmetry $\hat\Gamma$ in \emph{chiral SCs with strong SOCs}, in optical absorption spectroscopy and Raman spectroscopy. }
    \label{tab:chiral SC w/ SOC}
\end{table}

\begin{table}[h]
    \centering
    \begin{tabular}{|c|c|c|c|c|}
    \hline
         Parity of $\Delta(\mathbf{k})$ & $(\hat \Gamma^\prime)^2$ & Parity of $\epsilon^{\alpha\beta}\gamma^\dagger_{\mathbf{k},0,\alpha}\gamma^\dagger_{-\mathbf{k},0,\beta}$ & Im$\chi(2\Delta<\omega<\Delta+\Delta^\prime)$&$I_\text{Raman}(2\Delta<\omega<\Delta+\Delta^\prime)$\\
    \hline
        Even&+1& Even&0&Nonzero\\
        \hline
        Odd&-1&Odd&Nonzero&0\\
    \hline
    \end{tabular}
    \caption{Selection rules for the parity of pairing wavefunctions under inversion symmetry $\hat\Gamma$ in \emph{singlet SCs with no SOC} and hence $SU(2)$ symmetry, in optical absorption spectroscopy and Raman spectroscopy. }
    \label{tab:singlet SC w/o SOC}
\end{table}

\begin{table}[h]
    \centering
    \begin{tabular}{|c|c|c|c|c|}
    \hline
         Parity of $\Delta(\mathbf{k})$ & $(\hat C_{2,z}^\prime)^2$ & Parity of $\gamma^\dagger_{\mathbf{k},a}\gamma^\dagger_{-\mathbf{k},a}$ & Im$\chi(2\Delta<\omega<\Delta+\Delta^\prime)$&$I_\text{Raman}(2\Delta<\omega<\Delta+\Delta^\prime)$\\
    \hline
        Even&$-1$& Even&0&Nonzero\\
        \hline
        Odd&$+1$&Odd&Nonzero&0\\
    \hline
    \end{tabular}
    \caption{Selection rules for the parity of pairing wavefunctions under rotational symmetry $\hat C_{2,z}$ in \emph{2d chiral SCs with strong SOCs}, in optical absorption spectroscopy and Raman spectroscopy. }
    \label{tab:c2:chiral SC w/ SOC}
\end{table}
\twocolumngrid

\begin{figure}[h]
    \centering
    \includegraphics[width = \columnwidth]{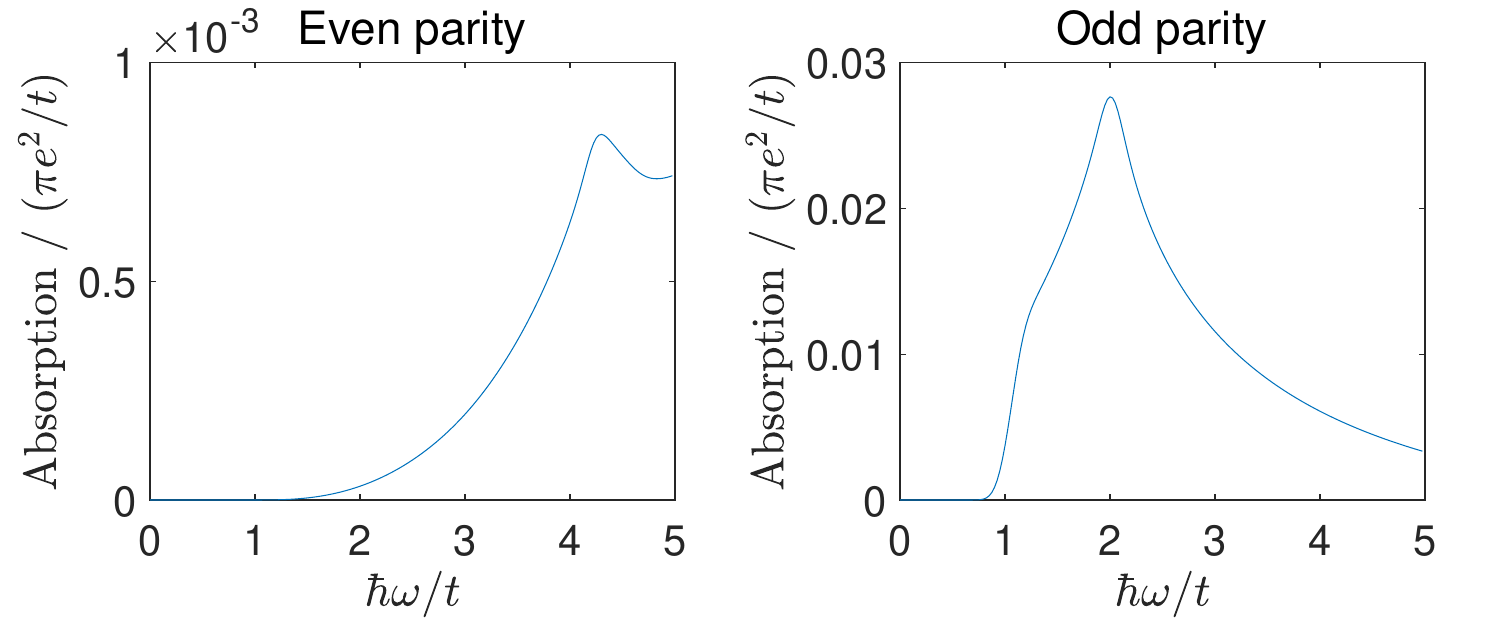}
    \caption{$T=0$ absorption spectra of two chiral SCs with opposite pairing parity under inversion symmetry, in a 2-band model of doped Weyl semimetals\cite{Yang2011,Cho2012} that breaks time reversal symmetry. (Left) $\Delta=0.5 t,~\Delta^\prime=3.5 t$; (Right) $\Delta=0,~\Delta^\prime=0.5 t$.}
    \label{fig:absorption:chiral}
\end{figure}

Below we demonstrate the absorption selection rules in two examples. First we study superconductivity in doped magnetic Weyl semimetals\cite{Yang2011,Cho2012} as an example of chiral SCs with strong SOCs. Specifically we consider two different pairing symmetries\cite{Supp} of zero-momentum Cooper pairs, one with even parity under inversion and another with odd parity, in the 2-band model of an inversion-symmetric Weyl semimetal that breaks time reversal symmetry\cite{Yang2011}. As shown in Fig. \ref{fig:absorption:chiral}, while the electronic contribution to light absorption remains finite at all frequency $\omega>2\Delta=0$ for an even-parity SC, it vanishes at a low frequency ($\omega<\Delta+\Delta^\prime$ in Fig. \ref{fig:absorption:chiral}) for an odd-parity SC. Here, for the even-parity SC, the energy gaps are $\Delta = 0.5 t, \Delta^\prime = 3.5 t$. Here, $t$ is the hopping integral defined in the appendix, which is the characteristic energy scale of our models.  The odd-parity SC state is gapless ($\Delta=0$), due to the nontrivial topological charge of the Weyl points\cite{Li2018a}. The energy gaps are $\Delta = 0, \Delta^\prime = 0.5 t$. Remarkably in the odd-parity SC, independent of the direction of light polarization $\mathbf{e}$\cite{Supp}, there is no optical absorption even above the thermodynamic gap $2\Delta$, and the associated "optical gap" is lower bounded by $\Delta_\text{op}\geq\Delta+\Delta^\prime$, consistent with selection rules in Table \ref{tab:chiral SC w/ SOC}. 

\begin{figure}[h]
    \centering
    \includegraphics[width = \columnwidth]{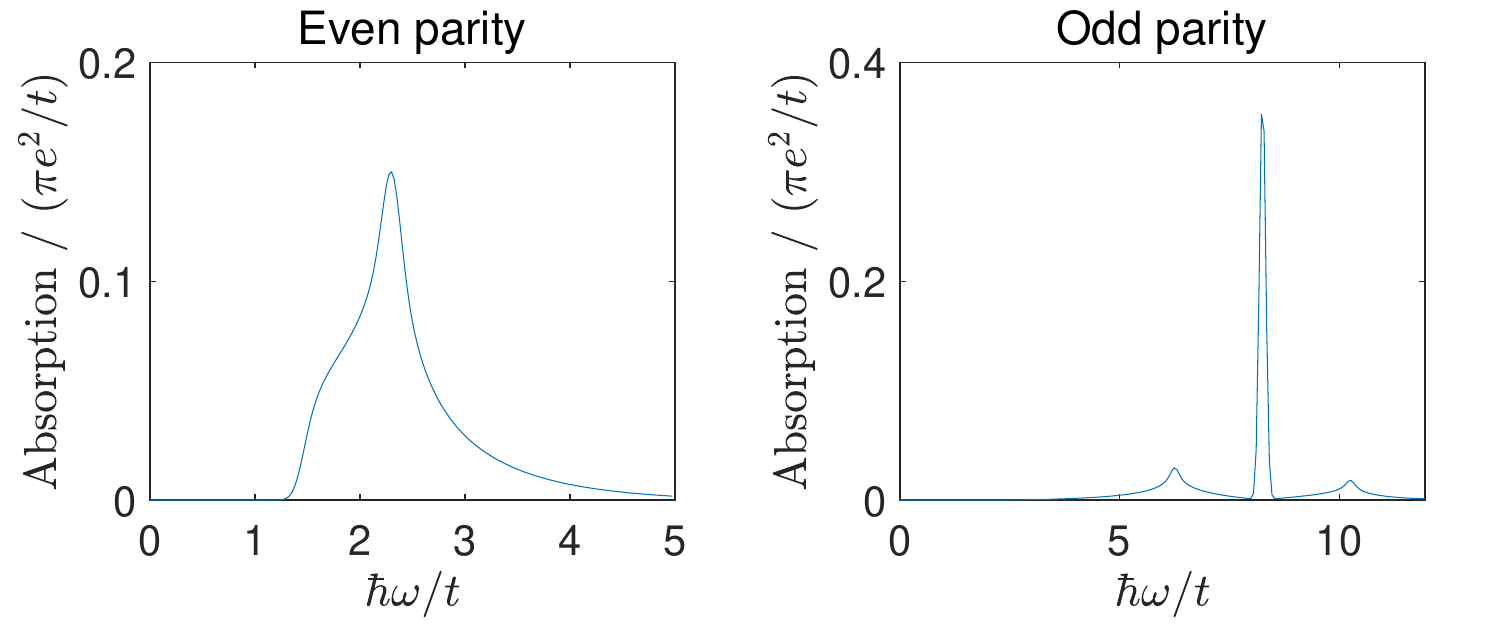}
    \caption{$T=0$ absorption spectra of two singlet SCs with opposite pairing parity under inversion symmetry, in the honeycomb lattice model of doped graphene. (Left) $\Delta=0.22 t,~\Delta^\prime=0.75 t$; (Right) $\Delta=1.1 t,~\Delta^\prime=4.2 t$.}
    \label{fig:absorption:singlet}
\end{figure}

Next we study spin-singlet superconductivity in doped graphene, as an example of singlet SCs with $SU(2)$ spin rotational symmetry. Specifically we consider two different pairing symmetries\cite{Supp} of zero center-of-mass momentum, in the honeycomb lattice model of graphene with a finite chemical potential. Energy gaps for even-parity case are $\Delta = 0.22 t, \Delta^\prime = 0.75 t$, and for odd-parity case are $\Delta = 1.1 t, \Delta^\prime = 4.2 t$. The absorption rate as a function of light frequency $\omega$ is shown in Fig. \ref{fig:absorption:singlet}. The odd-parity SC is gapless, and its absorption rate is nonzero for all frequency. In contrast, the absorption rate of the even-parity SC is identically zero above the bulk thermodynamic gap $2\Delta\approx 0.4 t$, until the light frequency reaches an ``optical gap'' of $\Delta_\text{op}\approx 1.2 t>\Delta+\Delta^\prime$. This demonstrates the selection rules for even-parity singlet SCs. 

The selection rules for optical absorption (and conductivity) in Table \ref{tab:chiral SC w/ SOC} and \ref{tab:singlet SC w/o SOC} have been previously discussed by Ref.\cite{Ahn2021} in the context of Altland-Zirnbauer classes\cite{Altland1997} of quadratic BdG Hamiltonians. Compared to Ref.\cite{Ahn2021} which is based on particle-hole symmetry of the quadratic BdG Hamiltonian, in the current work we derived the selection rules using the relation between fermion PSGs and SC pairing symmetry\cite{Yang2024,Shiozaki2022}, which generally applies to interacting fermions. In an interacting fermion system, since the Altland-Zirnbauer symmetry classes does not directly apply to a many-body interacting Hamiltonian, the symmetry group characterized by the fermion PSGs is required to properly describe the full symmetry of the superconducting phase. For example, in the case of global symmetries, class D corresponds to chiral superconductors w/o spin rotational symmetries (e.g. due to a strong spin-orbit coupling), class C corresponds to chiral singlet superconductors with $SU(2)$ spin rotational symmetries, while class CI corresponds to singlet superconductors with time reversal symmetry. As we have shown above, the parity of the pairing order parameter of these 3 symmetry types can be detected by selection rules, in the presence of inversion symmetry. Therefore our formulation enables a generalization of the single-particle results of Ref.\cite{Ahn2021} to correlated systems with well-defined Bogoliubov quasiparticles $\gamma_{{\bf k},a}$, in a way similar to how a fermi liquid generalizes a non-interacting fermi gas. Moreover, compared to Ref.\cite{Ahn2021}, our results point to a clear protocol to detect the SC pairing symmetry through a comparison of the optical gap and the thermodynamic gap $2\Delta$, which can be obtained by e.g. measuring the specific heat. Based on the above relation between fermion PSG and the SC pairing symmetry, below we further derive new selection rules in Raman spectroscopy.  

{\section{Raman spectroscopy}}

The Raman differential scattering cross section is given by the Fermi's golden rule\cite{Shastry1990,Shastry1991,Devereaux2007}:
\bea\notag \label{eq.raman}
&\frac{\partial^2 \sigma}{\partial \Omega \partial \omega_s}= \frac{e^4}{\hbar} \frac{\omega_s}{\omega_i}\\
& \sum_f|\expval{f|e_i^\alpha{e}_s^\beta M^{\alpha \beta}|i}|^2\delta(\omega+E_i-E_f)~~~
\eea
where $e_i$ and $e_s$ are the polarization vectors of incident and scattered light. $\omega_i,\omega_s$ are the frequencies of the incident and of the scattered light respectively, and $\omega=\omega_i-\omega_s$. $\hat M$ is Raman scattering operator given by
\bea\notag \label{eq.raman_M}
&\langle f|e_i^a {e}_f^b M^{\alpha \beta}| i\rangle={e}_{{i}}^a{{e}}_f^b\langle{f}|\sum_{{\mathbf{k},\alpha,\beta}}{c}_{{\mathbf{k}, \alpha}}^\dagger \frac{\partial^2 h_{\alpha,\beta}(\mathbf{k})}{\partial k^a \partial k^b}  {c}_{\mathbf{k}, \beta}|{{i}}\rangle\\
&+\sum_v\Big[\frac{\expval{ f|(\vec{j} \cdot \hat{e}_f)| v}\expval{v|(\vec{j} \cdot \hat{e}_i)| i}}{E_v-E_i-\omega_i}+\frac{\expval{f|(\vec{j} \cdot \hat{e}_i)| v}\expval{v|(\vec{j} \cdot {\hat{e}}_f)| i}}{E_v-E_i+\omega_s}\Big].
\eea
where $E_v$ labels the energy of eigenstate $\dket{v}$. Clearly $M$ operator is even under inversion symmetry. 

As a result, following the same logic as in the discussions of absorption spectroscopy, we can obtain the selection rules for Raman intensity $I_\text{Raman}(\omega)$ shown in the last column of Table \ref{tab:chiral SC w/ SOC}-\ref{tab:singlet SC w/o SOC}. Note that the inversion symmetry selection rules for Raman spectra is opposite to those for absorption spectra. In particular, for chiral SCs with strong SOCs, the Raman intensity vanishes even above the thermodynamic gap $2\Delta$ for an even-parity pairing wavefunction. 

\begin{figure}[h]
    \centering
    \includegraphics[width = \columnwidth]{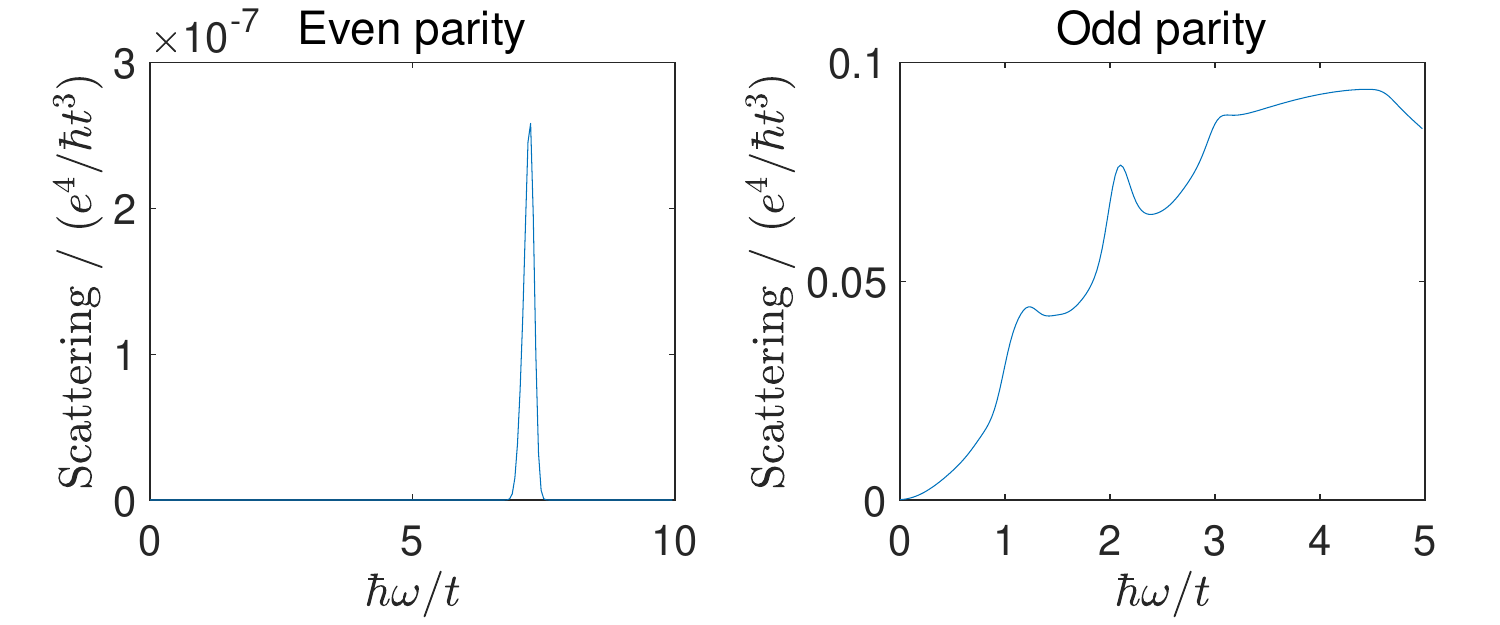}
    \caption{$T=0$ Raman spectra of two chiral SCs with opposite pairing parity under inversion symmetry, in a 2-band model of doped Weyl semimetals. (Left) $\Delta=0.5 t,~\Delta^\prime=3.5 t$; (Right) $\Delta=0 ,~\Delta^\prime=0.5 t$.}
    \label{fig:raman:chiral}
\end{figure} 

\begin{figure}[h]
    \centering
    \includegraphics[width = \columnwidth]{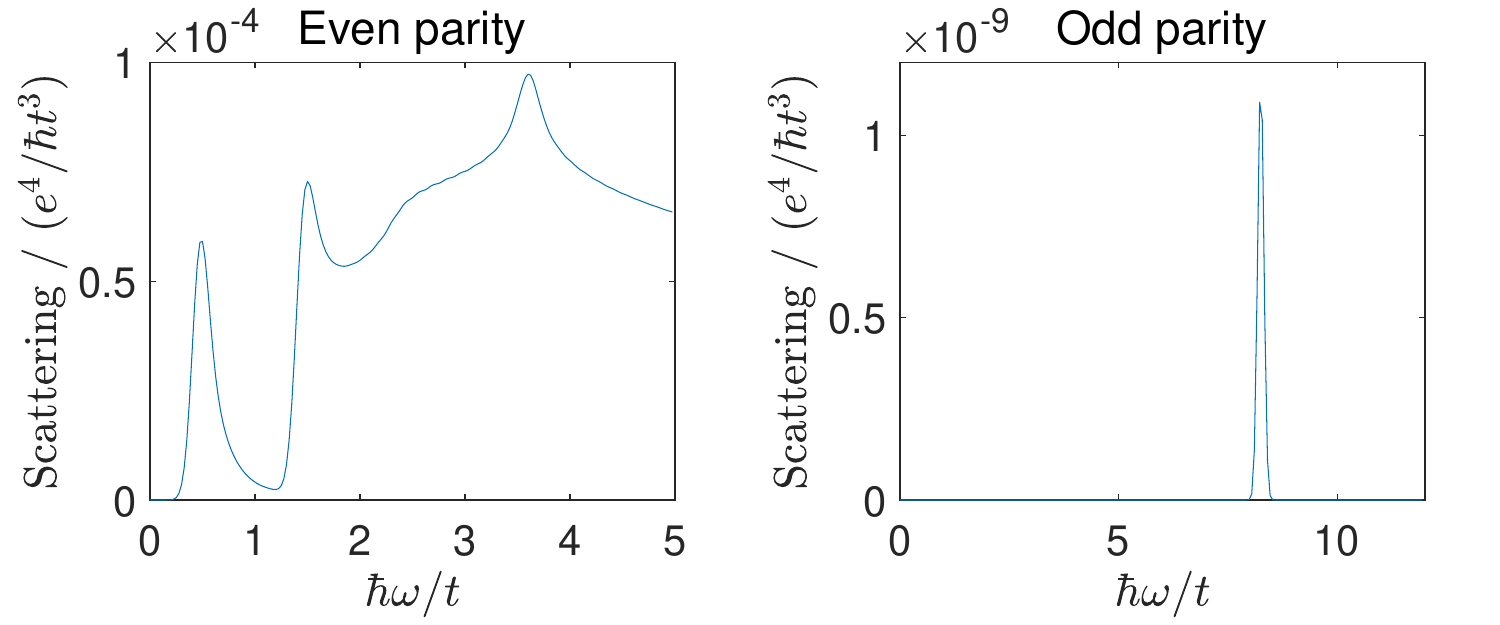}
    \caption{$T=0$ Raman spectra of two singlet SCs with opposite pairing parity under inversion symmetry, in the honeycomb lattice model of doped graphene. (Left) $\Delta=0.22 t,~\Delta^\prime=0.75 t$; (Right) $\Delta=1.1 t,~\Delta^\prime=4.2 t$.}
    \label{fig:raman:singlet}
\end{figure}

We also calculate Raman spectra for both our models. For the model of Weyl semimetals without $SU(2)$, we show results in Fig. \ref{fig:raman:chiral}. For even-parity case, the selection rule requires the scattering to be zero below $\Delta + \Delta^\prime$. For odd-parity case, there is no selection rule, and the spectrum is not zero at any frequency because the band is gapless ($\Delta = 0$).

The Raman spectra for SC on honeycomb lattice with $SU(2)$ symmetry is shown in Fig. \ref{fig:raman:singlet}. For even-parity case, the spectrum is not zero above energy gap $2\Delta = 0.44 t$ because there is no selection rule. For the odd-parity case, the sepctrum is zero between energy gap $2\Delta = 2.2 t$ and $\Delta + \Delta^\prime = 5.3 t$, as requested by selection rule.

{\section{$C_{2,z}$ symmetry in 2d superconductors}}
For a quasi-two-dimensional (2d) SC, the previous derivation of selection rules for inversion symmetry $\Gamma$ naturally generalizes to 2-fold rotation $C_{2,z}$ whose rotation axis is perpendicular to the 2d plane. There is one subtlety though, from nontrivial $C_{2,z}$ symmetry actions on electrons in the normal state: $\hat C_{2,z}^2=1$ for $SU(2)$-symmetric 2d systems with a weak SOC, or $\hat C_{2,z}^2=(-1)^{\hat F}$ for 2d systems with a strong SOC. In a 2d singlet SC with $SU(2)$ spin rotational symmetry, the 2-fold rotation $C_{2,z}$ simply reverses both $x$ and $y$ coordinates in the 2d space, in the same way as the inversion symmetry. This leads to the same set of selection rules in Table \ref{tab:singlet SC w/o SOC} for 2d singlet SCs with $C_{2,z}$ symmetry. On the other hand, for 2d chiral SCs with a strong SOC, due to the nontrivial $\hat C_{2,z}^2=(-1)^{\hat F}$ action in the normal state, the $C_{2,z}$ selection rules are opposite to those of inversion symmetry, as summarized in Table \ref{tab:c2:chiral SC w/ SOC}.

Below we briefly discuss one example where such selection rules apply. By analyzing a collection of existing experimental data, Ref.\cite{Lake2022} recently argued that the structure of SC pairing order parameters in magic-angle twisted bilayer graphene (MATBG) is almost uniquely determined except for its orbital parity under $C_{2,z}$ symmetry, which is either even ($d$-wave) or odd ($p$-wave). In the normal state of MATBG, $C_{2,z}^2=(-1)^{\hat F}$ arises from spontaneously generated SOC due to spin-valley locking\cite{Lake2022}, and therefore the selection rules in Table \ref{tab:c2:chiral SC w/ SOC} directly point to experimental signatures to determine the pairing symmetry of the chiral SC in MATBG\cite{Yang2024}.

{\emph{Discussions and Outlook}}~~~Using the correspondence between pairing symmetry and fermion PSG in SCs obtained recently\cite{Yang2024,Shiozaki2022}, we derived selection rules in optical absorption and Raman scattering spectroscopy for two classes of clean SCs: (i) chiral SCs with strong SOC and (ii) singlet SCs with weak SOC. In particular, the selection rules in case (i) can be applied to distinguish the inversion parity of pairing wavefunctions in many candidate materials of chiral superconductors\cite{Stewart2017}, such as SrRuO$_4$\cite{Maeno2012,Liu2015a}, UPt$_3$\cite{Avers2020} and UTe$_2$\cite{Aoki2022}. We derive a similar set of selection rules for 2-fold rotational symmetry $C_{2,z}$ in 2d SCs, and discuss how they can be applied to distinguish the proposed $d$-wave from $p$-wave SCs in MATBG\cite{Lake2022}. 

While this work focused on clean SCs, it is well known that optical absorption exists above the thermodynamic gap $2\Delta$ in dirty SCs\cite{Mattis1958,Leplae1983,Zimmermann1991}. Since impurities break the crystalline inversion or rotation symmetry, the selection rules derived here for clean SCs do not apply in the dirty limit, giving rise to nonzero optical and Raman responses above $2\Delta$. In presence of impurities, one natural question is, how does the optical responses in absorption and Raman scattering depend on the frequency $\omega$ above the thermodynamic gap $2\Delta$? For example, even- and odd-parity SCs may have different power-law dependence on $\omega-2\Delta$ in optical absorption and in Raman intensity. Another future direction is to generalize the current work to selection rules of optical and Raman responses in topological orders, e.g. $Z_2$ spin liquids which are related to a SC by gauging the fermion parity symmetry $Z_2^F$. We leave these questions for future studies. \\

\acknowledgements{We thank Seishiro Ono, Sayak Biswas and Mohit Randeria for helpful discussions and related collaborations. This work is supported by Center for Emergent Materials at The Ohio State University, a National Science Foundation (NSF) MRSEC through NSF Award No. DMR-2011876. YML acknowledges support by grant NSF PHY-1748958 to the Kavli Institute for Theoretical Physics (KITP), and NSF PHY-2210452 to the Aspen Center for Physics.}\\

\newpage

\onecolumngrid

\appendix


\section{Chiral superconductivity in a 2-band model of doped Weyl semimetal}

In this example, we demonstrate the phenomenon of selection rule by studying superconductivity in a doped Weyl semi-metal. In this example of a doped Weyl semimetal, time reversal symmetry is broken, while inversion and $C_4$ rotational symmetry along $z$ axis are preserved. To be more specific, we consider nearest neighbour terms in a $3D$ cubic lattice model with $2$ orbitals on each site, and the Hamiltonian\cite{Yang2011,Cho2012} in the momentum space is:
\begin{equation}
    h_0(\mathbf{k}) = t\ sin(k_x) \sigma^x + t\ sin(k_y) \sigma^ y + m(2 - cos(k_x) -cos(k_y)) \sigma^z + t_z(cos(k_z) - cos(Q)) \sigma^z + \mu \sigma^0
\end{equation}
Here, $t(k)$ means it is hopping part of the Hamiltonian, $t, m, t_z, \mu$ are real numbers. Two dimensions of matrix $\sigma$ represents $2$ orbitals. This Hamiltonian generally has Weyl nodes at $\mathbf{k} = (0, 0, \pm Q)$.

Inversion and $C_4$ rotational symmetries act on fermions in the following way:
\begin{equation}
    \begin{aligned}
        I: & \quad c_{s, \mathbf{k}} \rightarrow \sigma^z_{s, s'}c_{s', -\mathbf{k}}\\
        C_4: & \quad c_{s, (k_x, k_y, k_z)} \rightarrow S_{s, s'} c_{s', (k_y, -k_x, kz)}
    \end{aligned}
\end{equation}
where $\sigma^z$ and $S = \frac{1}{\sqrt{2}} (\sigma^0 + i \sigma^z)$ is action of $C_4$ on orbitals. Inversion and $C_4$ rotation act both on space (momentum is changed) and spin (matrix $\sigma$) degrees of freedoms.
In the form of matrix, we can write
\begin{equation}
    \begin{aligned}
        I : & \quad h_0(-\mathbf{k}) \rightarrow \sigma^z h_0(\mathbf{k}) \sigma^z \\
        C_4 : & \quad h_0(-k_y, k_x, k_z) \rightarrow S^\dagger h_0 (k_x, k_y, k_z) S
    \end{aligned}
\end{equation}

To have superconductivity, we add pairing term $\Delta^\dagger_{s_1, s_2}(k)\ c_{s_1, -k} c_{s_2, k} + h.c.$. Inversion and rotational symmetry acting on the pairing terms in the following way:
\begin{equation}
    \begin{aligned}
        I : & \quad \Delta(-\mathbf{k})  \rightarrow \sigma^z \Delta(\mathbf{k}) \sigma^z \\
        C_4 : & \quad \Delta(-k_y, k_x, k_z) \rightarrow S^\dagger \Delta(k_x, k_y, k_z) S^*
    \end{aligned}
\end{equation}
Under this transformation, different pairing terms have different quantum number and they are listed in Table \ref{table.pairing_term}.

\begin{table}[h]
    \centering
    \begin{tabular}{|c|c|c|c|c|c|}
    \hline
        $\Delta(\mathbf{k})$   &  $ I$ & $C_4$ & $I^{\prime2}$ & $C_4^{\prime 4}$ & $I^\prime C_4^\prime I^{\prime -1} C_4^{\prime -1}$\\
    \hline
        $\cos{k_z} i\sigma^y$ & -1 & 1 & -1 & -1 & 1\\
    \hline
        $\sin{k_z} \sigma ^ x$ & 1 & 1 & 1 & -1 & 1\\
    \hline
        $\sin{k_z} (\sigma^0 + \sigma^z) $ & -1 & -i & -1 & 1 & 1\\
    \hline
        $\sin{k_z} (\sigma^0 - \sigma^z) $ & -1 & i & -1 & 1 & 1\\
    \hline
        $(\cos{k_x} + \cos{k_y}) i\sigma^y$ & -1 & 1 & -1 & -1 & 1\\
    \hline
        $(\cos{k_x} - \cos{k_y}) i\sigma^y$ & -1 & -1 & -1 & -1 & 1\\
    \hline
        $(\sin{k_x} + i \sin{k_y}) \sigma ^ x$ & 1 & -i & 1 & 1 & 1\\
    \hline
        $(\sin{k_x} - i\sin{k_y}) \sigma ^ x$ & 1 & i & 1 & 1 & 1\\
    \hline
        $(\sin{k_x} + i \sin{k_y}) (\sigma^0 + \sigma^z)$ & -1 & -1 & -1 & -1 & 1\\
    \hline
        $(\sin{k_x} - i\sin{k_y}) (\sigma^0 + \sigma^z)$ & -1 & 1 & -1 & -1 & 1\\
    \hline
        $(\sin{k_x} + i \sin{k_y}) (\sigma^0 - \sigma^z)$ & -1 & 1 & -1 & -1 & 1\\
    \hline
        $(\sin{k_x} - i\sin{k_y}) (\sigma^0 - \sigma^z)$ & -1 & -1 & -1 & -1 & 1\\
    \hline
    \end{tabular}
    \caption{For each pairing term, its quantum number under operators $I$ and $C_4$ are listed. The true symmetry $I^\prime, C_4^\prime$ have different symmetry fractionalization classes, the corresponding quantum number $I^{\prime2}, C_4^{\prime4}, I^\prime C_4^\prime I^{\prime-1} C_4^{\prime-1}$ are listed.}
    \label{table.pairing_term}
\end{table}

In this table $I$ and $C_4$ are the operators defined before. We show the quantum number of each term in the table. However, as discussed in the main text, symmetries of the BdG Hamiltonian are not $I$ and $C_4$, but different by a phase operator  $\exp{(i \Phi \hat{F})}$. Here $\hat{F}$ is fermion particle number operator. This phase operator transforms $\Delta$ as
\begin{equation}
    \Delta(\mathbf{k}) \rightarrow \Delta(\mathbf{k}) \exp{(i 2 \Phi)}
\end{equation}
Note that this phase operator does not influence hopping terms.
When pairing terms are not invariant under $I$ and $C_4$, we redefine $I^\prime = I \exp{(i \Phi \hat{F})}$ and $C_4^\prime = C_4 \exp{(i \Phi^\prime \hat{F})}$ so that $I^\prime$ and  $C_4^\prime$ are explicit symmetry of pairing terms. Then inversion symmetry $I^\prime$ and rotational symmetry $C_4^\prime$ are still preserved for the system. Different choices of terms lead to different symmetries fractionalizations characterized by $I^{\prime 2}$, $C_4^{\prime 4}$ and $I^\prime C_4^\prime I^{\prime-1} C_4^{\prime -1}$. We calculate these number for each pairing term and list them in Table \ref{table.pairing_term}.

The BdG Hamiltonian in the Nambu basis $\psi_\mathbf{k} = (c_{{\mathbf{k}}, \uparrow}, c_{{\mathbf{k}}, \downarrow}, c_{-{\mathbf{k}}, \uparrow}^\dagger, c_{-{\mathbf{k}}, \downarrow}^\dagger)^T$ is written as
\begin{equation}\label{eq.Hk_operator}
    \hat{H}_\text{BdG} = \frac{1}{2}\sum_{\mathbf{k}} \psi_{\mathbf{k}}^\dagger H_{\mathbf{k}} \psi_{\mathbf{k}}
\end{equation}
where 
\begin{equation}\label{eq.Hk}
    H_{\mathbf{k}} =
    \begin{bmatrix}
        h_0({\mathbf{k}}) &  \Delta({\mathbf{k}})\\
        \Delta^\dagger({\mathbf{k}}) &  -h_0^T(-{\mathbf{k}})\\
    \end{bmatrix}
\end{equation}
We have suppressed the orbital index $a$ in the fermion operator $c_{{\bf k},a,\uparrow/\downarrow}$. In other words, we use $c_{{\mathbf{k}}, \uparrow}$ to denote a spinor $c_{{\mathbf{k}}, \uparrow}=(c_{{\mathbf{k}},1,\uparrow}\cdots,c_{{\mathbf{k}},N,\uparrow})$ where $N$ is the number of bands in the system. 

Due to the particle-hole symmetry, the eigenvectors $u_{{\bf k},a}$ of above BdG Hamiltonian matrix $H_{\bf k}$ at momenta $\pm{\bf k}$ are related by:
\bea
&u_{-{\bf k},a}=\tau_x u_{{\bf k},a}^\ast,\\
&H_{\bf k}u_{{\bf k},a}=E_a({\bf k})u_{{\bf k},a},~~~H_{-{\bf k}}u_{-{\bf k},a}=-E_a({\bf k})u_{-{\bf k},a}.
\eea
where $(\tau_x,\tau_y,\tau_z)$ are Pauli matrices for the Nambu index. Therefore the eigenvectors of BdG Hamiltonian $H_{\bf k}$ include both ``quasiparticles'' $u_{{\bf k},a}$ with eigenvalue $E_a({\bf k})\geq0$ and ``quasiholes'' $\tau_x u^\ast_{-{\bf k},b}$ with eigenvalue $-E_b(-{\bf k})\leq0$. Their associated eigenmode operators are 
\bea
\gamma^\dagger_{{\bf k},a}=\sum_\alpha u_{{\bf k},a,\alpha}\psi^\dagger_{{\bf k},\alpha};~~~\gamma_{-{\bf k},b}=\sum_\alpha u^\ast_{-{\bf k},b,\alpha}\psi_{-{\bf k},\alpha}.
\eea
and the BdG Hamiltonain is diagonalized in the following form:
\bea\label{BdG diagonal form}
\hat H_\text{BdG}=\frac12\sum_{{\bf k},a}\Big[E_a({\bf k})\gamma^\dagger_{{\bf k},a}\gamma_{{\bf k},a}-E_a(-{\bf k})\gamma_{-{\bf k},a}\gamma^\dagger_{-{\bf k},a}\Big]
\eea

Next, to study optical spectrum of this model, we need to first obtain the electric current operator. The particle current operator can be calculated as derivative of hopping part of Hamiltonian. 
\begin{equation}\label{eq.current}
    \mathbf{\hat{j}} = \sum_{\mathbf{k}} (c_{{\mathbf{k}}, \uparrow}^\dagger, c_{{\mathbf{k}}, \downarrow}^\dagger) \  \mathbf{j} (\mathbf{k}) (c_{{\mathbf{k}}, \uparrow}, c_{{\mathbf{k}}, \downarrow}) ^ T
\end{equation}
with 
\bea
\mathbf{j} (\mathbf{k}) = \frac{\partial h_0({\mathbf{k}})}{\partial \mathbf{k}} 
\eea
Writing the current operator in the BdG basis $\psi_\mathbf{k} = (c_{{\mathbf{k}}, \uparrow}, c_{{\mathbf{k}}, \downarrow}, c_{-{\mathbf{k}}, \uparrow}^\dagger, c_{-{\mathbf{k}}, \downarrow}^\dagger)^T$, we have 
\bea
\mathbf{J}(\mathbf{k}) = 
    \begin{bmatrix}
        \frac{\partial h_0}{\partial \mathbf{k}} ({\mathbf{k}})&  0\\
        0 &  - \frac{\partial h_0^T}{\partial \mathbf{k}}(-\mathbf{k})
    \end{bmatrix}
\eea
and the current operator is
\bea
    \mathbf{\hat{j}} = \frac{1}{2}\sum_{\mathbf{k}} \psi_{\mathbf{k}}^\dagger \mathbf{J}(\mathbf{k}) \psi_{\mathbf{k}}
\eea

The electric current is odd under inversion. Current in $z$ direction $\hat{j}_z$ is invariant under $C_4^\prime$ and $\hat{j}_x + i \hat{j}_y$ and $\hat{j}_x - i \hat{j}_y$ are two irreducible representation that carries phase $-i, i$ under $C_4^\prime$ rotation. Specifically,
\begin{equation}\label{eq.polarization}
    \begin{aligned}
        I^\prime (\hat{j}_x + i \hat{j}_y) I ^ {\prime-1} &= - (\hat{j}_x + i \hat{j}_y)\\
        I^\prime (\hat{j}_x - i \hat{j}_y) I ^ {\prime-1} &= - (\hat{j}_x - i \hat{j}_y)\\
        I^\prime \hat{j}_z I^{\prime-1} &= - \hat{j}_z\\
        C_4^\prime (\hat{j}_x + i \hat{j}_y) C_4 ^ {\prime-1} &= -i (\hat{j}_x + i \hat{j}_y)\\
        C_4^\prime (\hat{j}_x - i \hat{j}_y) C_4 ^ {\prime-1} &= i (\hat{j}_x - i \hat{j}_y)\\
        C_4^\prime \hat{j}_z C_4^{\prime-1} &= - \hat{j}_z
    \end{aligned}
\end{equation}

We calculate current operator from \eqref{eq.current} as follows:
\begin{equation}
    \begin{aligned}
        j_x (\mathbf{k})=& t\cos{k_x} \sigma^x + m \sin{k_x} \sigma^z\\
        j_y (\mathbf{k}) =& t\cos{k_y} \sigma^y + m \sin{k_y} \sigma^z\\
        j_z(\mathbf{k}) =& -t_z \sin{k_z} \sigma^z 
    \end{aligned}
\end{equation}

We show two examples.

Case 1: $I^{\prime2} = 1, C_4^{\prime4} = -1$. In this case $I^\prime = I, C_4^\prime = C_4$ are not changed. There is the only one pairing term invariant according to Table. \ref{table.pairing_term}:
\begin{equation}
     \Delta(\mathbf{k}) = \Delta_1 \sin{k_z} \sigma^x
\end{equation}
Here $\Delta_1$ is a complex number.
Parameters are set as $Q = \pi / 2, \ \mu = 3.5t, \ m = 0.5t, \ t_z = t, \ \Delta_1 = t$.

Case 2: $I^{\prime2} = -1, C_4^{\prime4} = -1$. And $I^\prime = I \exp{(i \pi \hat{F} / 2 )}$, $C_4^\prime = C_4$. In this case, all pairing terms with phase $-1, 1$ under transformation of $I, C_4$ are invariant under $I^\prime$ and $C_4^\prime$. According to Table. \ref{table.pairing_term}, pairing terms in the Hamiltonian are
\begin{equation}
    \Delta(\mathbf{k})= \Delta_1^\prime\cos{k_z} i\sigma^y + \Delta_2^\prime (\cos{k_x} + \cos{k_y}) i\sigma^y + \Delta_3^\prime (\sin{k_x} + i \sin{k_y}) (\sigma^0 - \sigma^z) + \Delta_4^\prime (\sin{k_x} - i\sin{k_y}) (\sigma^0 + \sigma^z)
\end{equation}

Parameters are set as $Q = \pi / 2, \ \mu = 0.3 t, \ m = t, \ t_z = t, \ \Delta_i^\prime = 0.2 t$.

\begin{figure}[h]
    \centering
    \includegraphics[width = 0.48 \columnwidth]{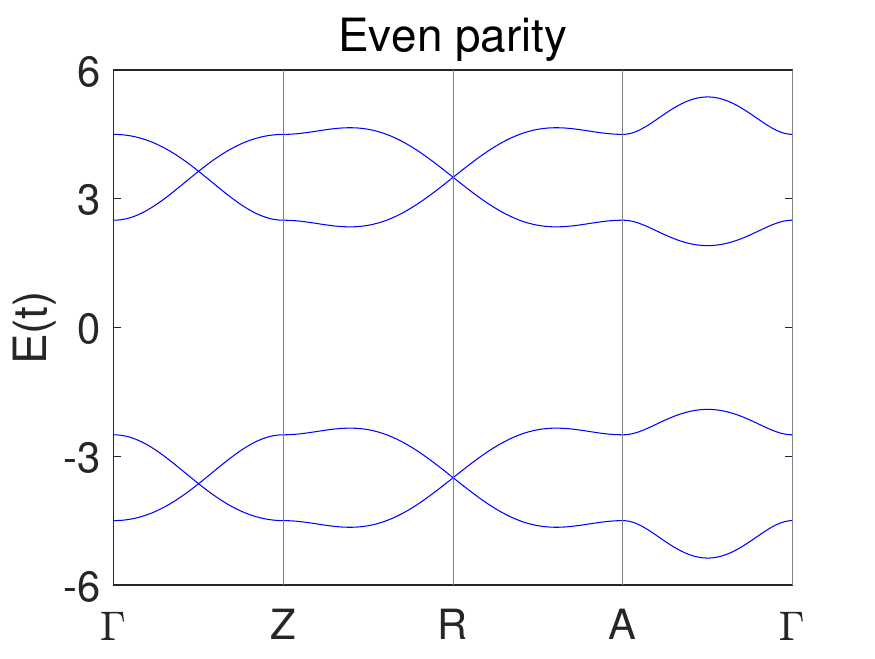}
    \includegraphics[width = 0.48 \columnwidth]{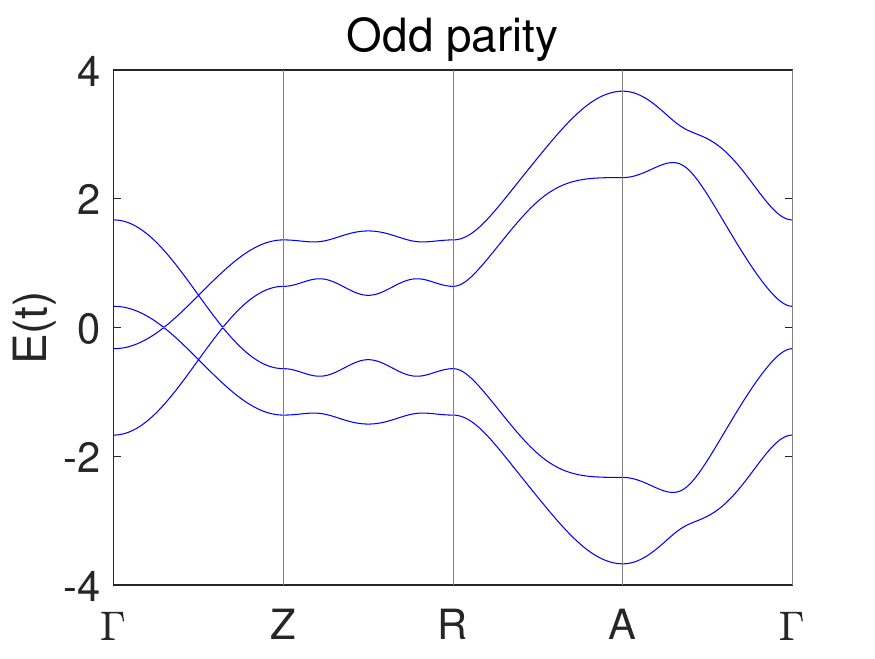}
    \caption{Energy band of electrons. Special points in Brillioun Zone are $\Gamma : (0, 0, 0), \ Z:(0, 0, \pi),\ R: (0, \pi, \pi), \ A:(\pi, \pi, \pi)$. Gapless point between $\Gamma$ and $Z$ is part of a sphere of gapless points. For odd-parity case, the energy gap of first band is $\Delta = 0.5 t$ and the energy gap of the second band is $\Delta^\prime = 3.5 t$. For even-parity case, the energy gap of first band is $0$ and the energy gap of the second band is $\Delta^\prime = 0.5 t$.}
    \label{fig:sqr_band}
\end{figure}

\begin{figure}[h]
    \centering
    \includegraphics[width = 0.48 \columnwidth]{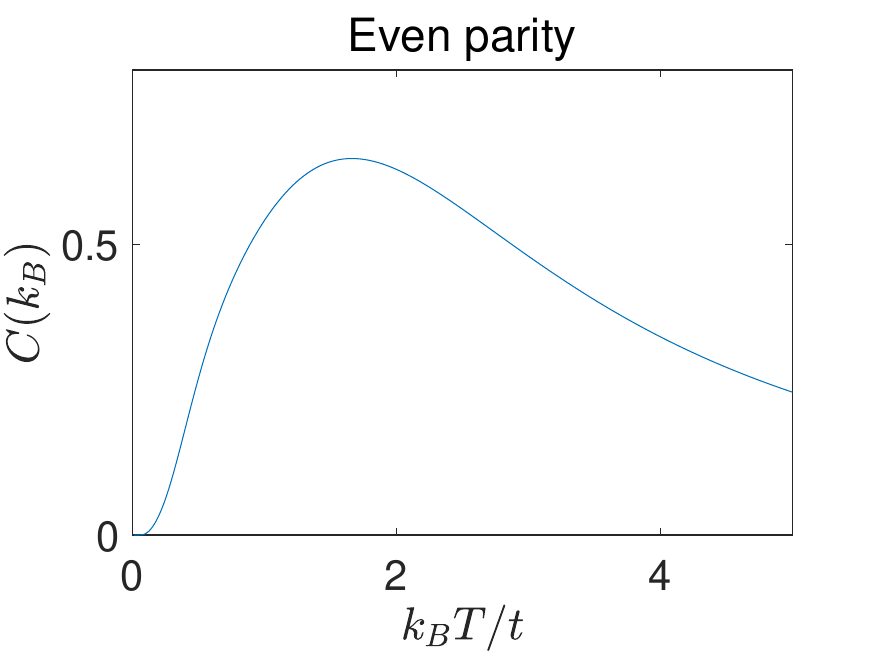}
    \includegraphics[width = 0.48 \columnwidth]{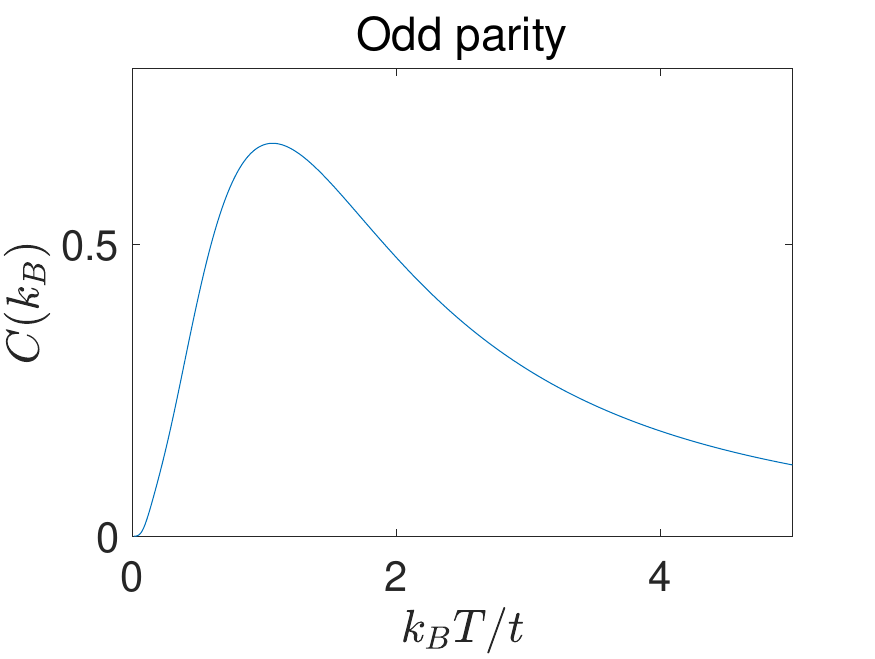}
    \caption{Specific heat versus temperature T. Even-parity case is gapped with $\Delta=0.5 t$ and specific heat is exponential function of $1/T$ at low temperature. Odd-parity case has gapless momentum points and specific heat is proportional to $T^3$ at low temperature.}
    \label{fig:sqr_heat_capacity}
\end{figure}

We show gapless electron bands in Fig. \ref{fig:sqr_band} and specific heat in Fig. \ref{fig:sqr_heat_capacity}. Specific heat is calculated by 
\bea\label{eq.specific_heat}
C = \frac{1}{V} \sum_{i} \frac{\epsilon_i ^ 2 \exp{(\epsilon_i/T)}}{T^2 \left(1 + \exp{(\epsilon_i / T)} \right) ^ 2}
\eea
where $\epsilon_i$ sums over all positive eigenvalues of the BdG Hamiltonian, index $i$ including momentum index $\mathbf{k}$ and band index $a$. $V$ is the number of unit cells. Even-parity case is gapped and heat capacity is close to exponential function $\exp{(-\Delta/T)}$ at low temperature. Odd-parity case has gapless points and heat capacity scales as $ T^3$ at low temperature.

\begin{figure}[h]
    \centering
    \includegraphics[width = 0.48 \columnwidth]{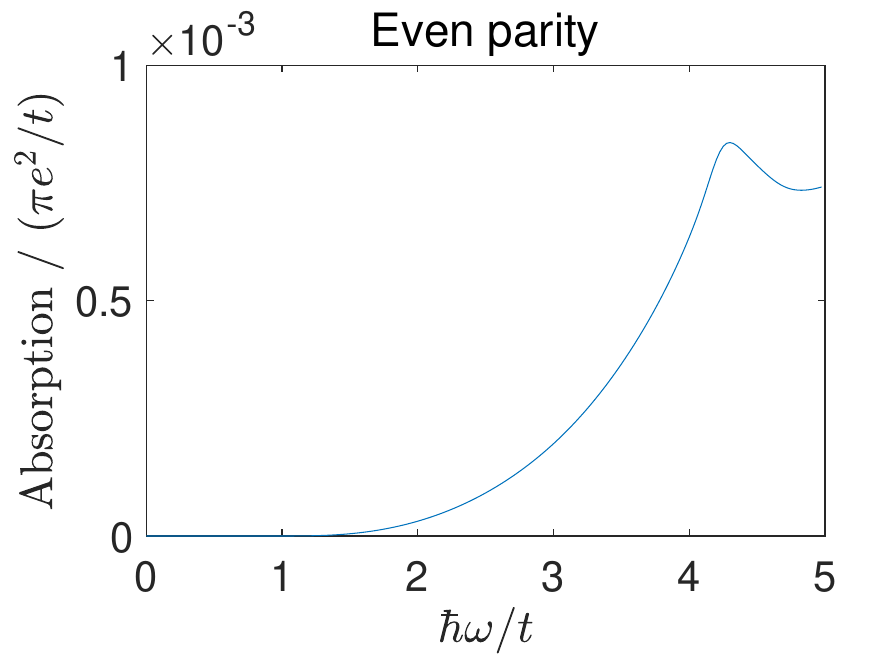}
    \includegraphics[width = 0.48 \columnwidth]{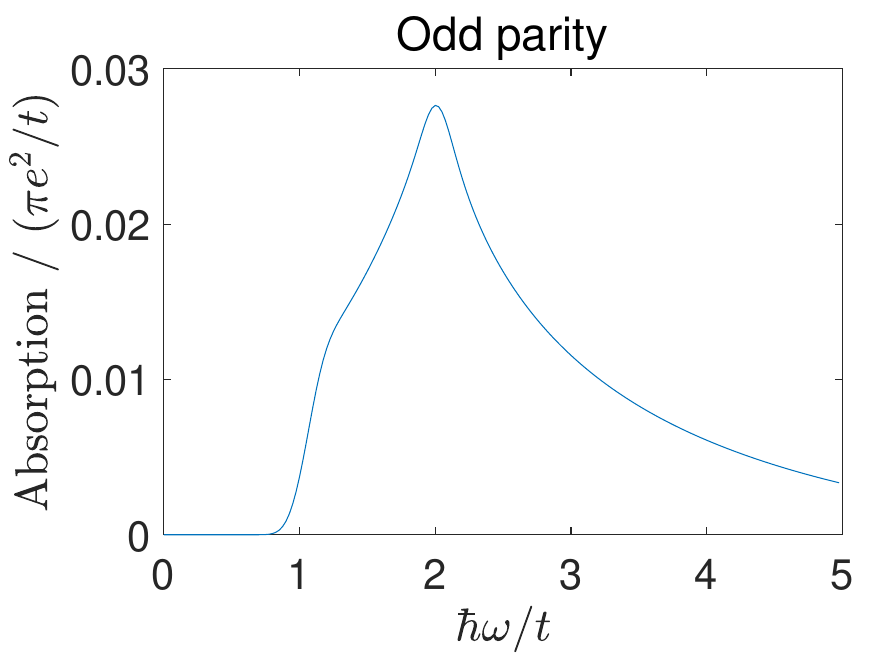}\\
    \includegraphics[width = 0.48 \columnwidth]{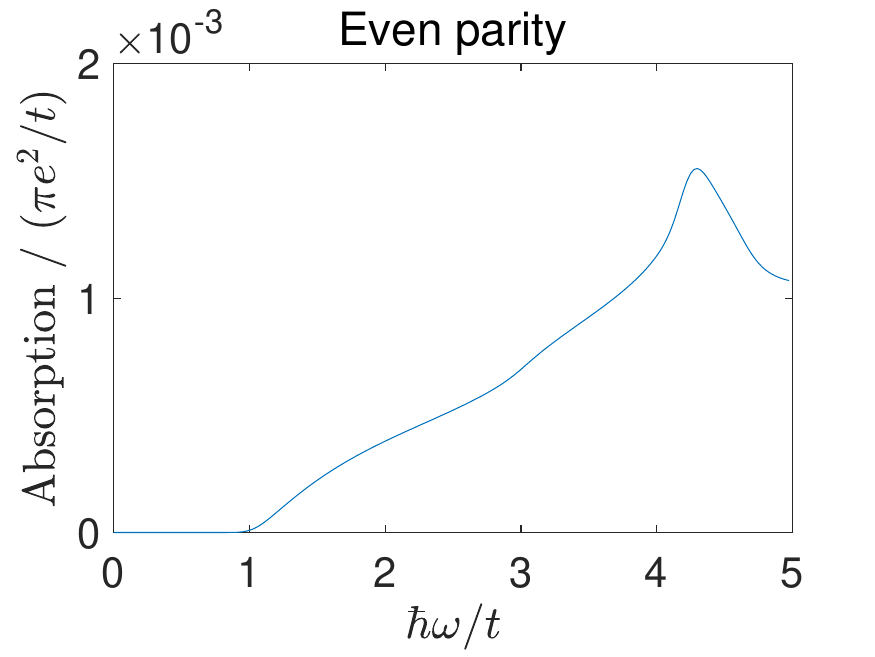}
    \includegraphics[width = 0.48 \columnwidth]{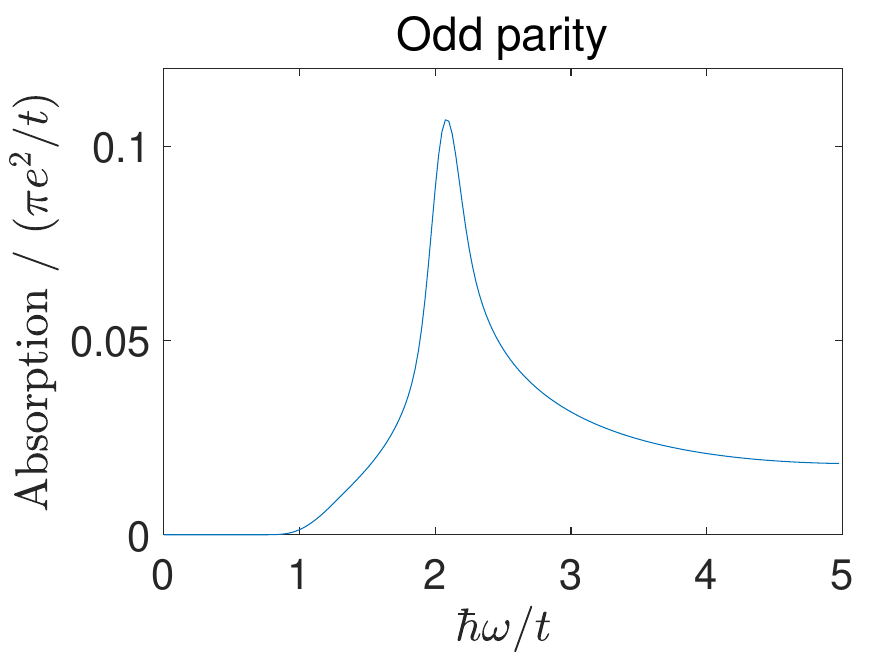}
    \caption{Absorption spectrum are intensity I versus photon energy $\hbar\omega$. Spectrum is corresponding to models with $I^2 = 1$ and polarization in $z$ direction (upper left) and $x$ direction (bottom left). In comparison, we show the spectrum of model with $I^2 = -1$ and polarization in $z$ direction (upper right) and $x$ direction (bottom right).}
    \label{fig:spectrum}
\end{figure}

We can calculate absorption spectrum exactly using Eq. \ref{eq.absorption}. For mean-field Hamiltonian, $|f\rangle$ can only be two particle states. At zero temperature, optical absorption rate is calculated by the following equations
\begin{equation}\label{eq.spectrum}
\begin{aligned}
    \omega~\text{Im}\chi(T=0,\omega)= &\frac{\pi e^2}{2\hbar \omega} \sum_{f} |\langle 0|\mathbf{e} \cdot \mathbf{\hat{j}}| f \rangle | ^ 2 \delta \left( \omega - E_f + E_0 \right) \\
    =& \frac{\pi e^2}{2\hbar \omega} \sum_{\mathbf{k}, a, b}  |\langle 0| \mathbf{e} \cdot \mathbf{\hat{j}} \gamma^\dagger_{\mathbf{k}, a} \gamma^\dagger_{-\mathbf{k}, b}| 0\rangle | ^ 2 \delta \left( \omega - E_a(\mathbf{k}) - E_b(-\mathbf{k}) \right) \\
    = & \frac{\pi e^2}{2\hbar \omega} \sum_{\mathbf{k}, a, b, \alpha, \beta} | \mathbf{e} \cdot \mathbf{J}^{\alpha, \beta}(\mathbf{k}) \langle 0| \psi_{\mathbf{k}, \alpha}^\dagger \  \psi_{\mathbf{k}, \beta} \gamma^\dagger_{\mathbf{k}, a} \gamma^\dagger_{-\mathbf{k}, b}| 0\rangle | ^ 2 \delta \left( \omega - E_a(\mathbf{k}) - E_b(-\mathbf{k}) \right) \\
    = &\frac{\pi e^2}{2\hbar \omega} \sum_{\mathbf{k}, a, b, \alpha, \beta} | \mathbf{e} \cdot \mathbf{J}^{\alpha, \beta}(\mathbf{k}) \langle 0| \psi_{\mathbf{k}, \alpha}^\dagger \gamma^\dagger_{-\mathbf{k}, b} |0\rangle \langle 0 |  \psi_{\mathbf{k}, \beta} \gamma^\dagger_{\mathbf{k}, a} | 0\rangle | ^ 2 \delta \left( \omega - E_a(\mathbf{k}) - E_b(-\mathbf{k}) \right)
\end{aligned}
\end{equation}
Here, we apply the general formula on the first line onto BdG system. Since our current operators are only up to quadratic order of fermion operators, we only excite two quasi-particles at the same time. So, final state is specified as $ \gamma^\dagger_{\mathbf{k}, a} \gamma^\dagger_{-\mathbf{k}, b} |0\rangle$ on the second line. $a, b$ are the labels of bands of positive energy and $\alpha, \beta$ are the labels of basis of BdG Hamiltonian. In the last line, we use Wick's theorem and this is the only contraction consistent with quantum number momentum. Since we have obtained the eigenstates of the Hamiltonian by diagonalization as
\bea
\gamma_{\mathbf{k}, a}^\dagger = \sum_{\alpha} u_{\mathbf{k},a,\alpha} \psi_{\mathbf{k}, \alpha}^\dagger
\eea
where $u_{\mathbf{k}, a, \alpha}$ is the $a$-th eigenvector of BdG Hamiltonian (\ref{eq.Hk}) at momentum ${\bf k}$, we can express the absorption rate more explicitly,
\bea\label{eq.spectrum_1}
 \omega ~\chi^{\prime\prime}(T=0,\omega) = \frac{\pi e^2}{2\hbar \omega} \sum_{a, b, \alpha, \beta}| u_{\mathbf{k}, b, \alpha} \mathbf{e} \cdot \mathbf{J}^{\alpha, \beta}(\mathbf{k}) u_{\mathbf{k}, a, \beta}^*|^2 \delta(\omega - E_a(\mathbf{k}) -E_b(-\mathbf{k}))
\eea
Here, index $a$ is summed over bands of positive energy $E_a(\mathbf{k})$, and index $b$ is summed over bands of negative energy $E_b(\mathbf{k})$.

In Fig. \ref{fig:spectrum}, we show absorption spectrum of our model under a polarized beam of light. Figures on left and right represent two models and they reveal stark contrasts in the spectrum. Figures on the left ( $I^{\prime2} = 1$ ) is normal, with non-zero amplitude above energy gap $2\Delta \approx  t$. In contrast, spectrum on the right ($I^{\prime2} = -1$) is a flat line of zero at low energy above energy gap $2\Delta = 0$, only grow to finite above energy threshold $\Delta + \Delta^\prime \approx 0.5 t$, indicating selection rule of $I^{\prime 2} = -1$. One detail is the spectum is not zero even with $\omega$ a little bit higher than $0.5 t$. That is because the band bottom of first and second band does not occur at the same momentum. This differences underscore the importance of inversion symmetry fractionalization on absorption spectrum of superconductors.

\begin{figure}[h]
    \centering
    \includegraphics[width = 0.48 \columnwidth]{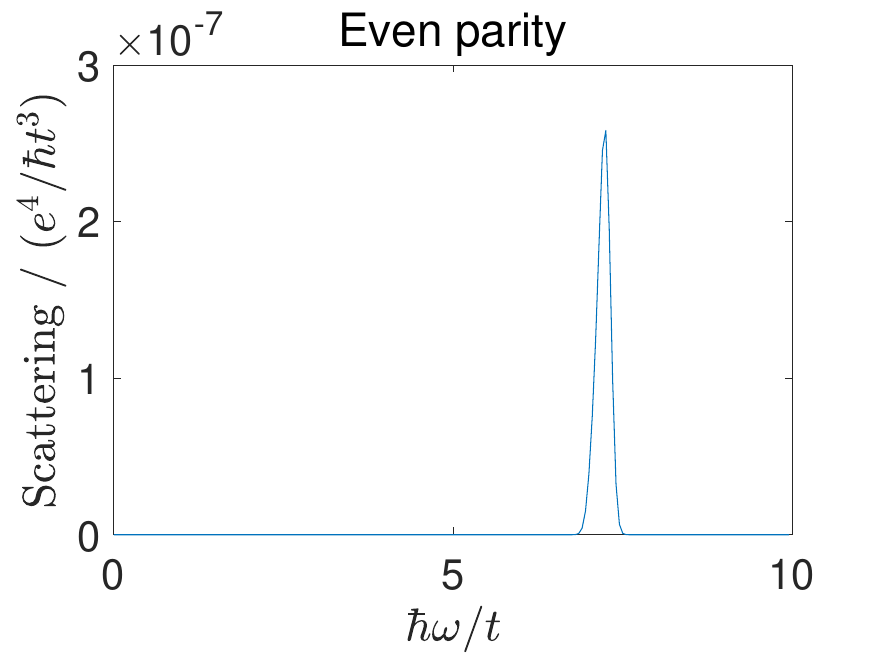}
    \includegraphics[width = 0.48 \columnwidth]{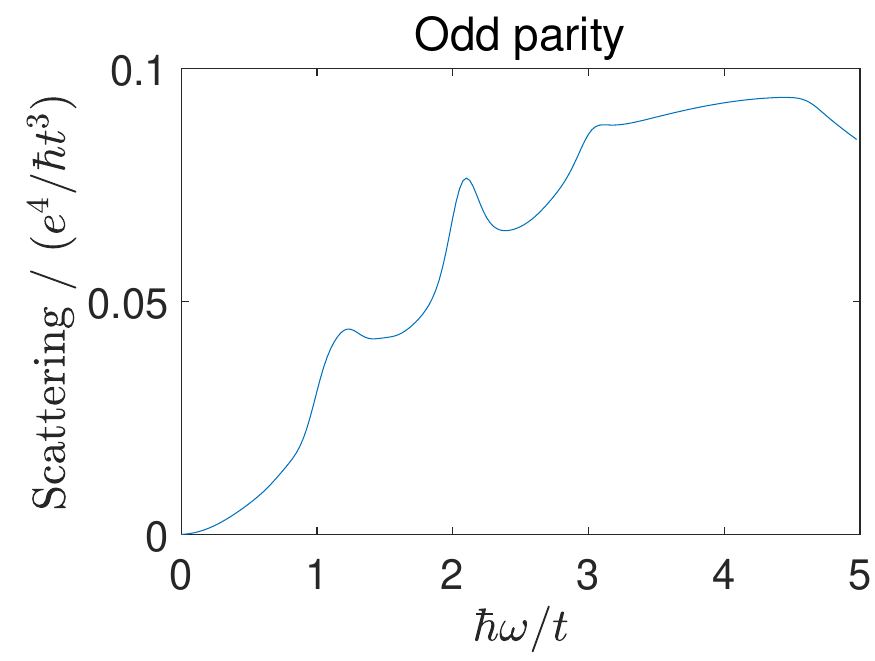}
    \caption{Differential scattering cross section of both incident and scattered light polarized in $x$ direction. Figures are corresponding to $I^2 = 1$ (left) and $I^2 = -1$ (right).}
    \label{fig:raman}
\end{figure}

We also show Raman spectrum of both models in Fig. \ref{fig:raman}. We calculate Raman differential scattering cross section per volume from Eq. \ref{eq.raman}. 
\bea\notag 
\frac{1}{V}\frac{\partial^2 \sigma}{\partial \Omega \partial \omega_s}= \frac{e^4}{\hbar V} \frac{\omega_s}{\omega_i} \sum_f|\expval{f|e_i^\alpha{e}_s^\beta M^{\alpha \beta}|i}|^2\delta(\omega+E_i-E_f)~~~
\eea
where $V$ is the number of unit cells. This calculation is more complex, but can still be done accurately for mean-field theory. We make use of Eq. \ref{eq.raman_M} and calculate expectation value for excitation of two particle states. This calculation is done by Wick's theorem similarly to the calculation of absorption spectrum. It is worth mentioning that four-particle states $|f\rangle$ has scattering amplitude $M$ zero and do not contribute to raman scattering in mean-field calculation.  Further details of this calculation refers to  papers \cite{Shastry1990,Shastry1991,Devereaux2007}.  

Raman spectrum has selection rule for even-parity case, which is opposite to the selection rule of absorption spectrum. In the even-parity case, Raman scattering is zero below $\Delta + \Delta ^\prime \approx 4 t$, because of the selection rule. In the odd-parity case, Raman scattering is not zero at any frequency because the band is gapless.

\section{Singlet superconductivity in doped graphene}

Our second example is a 2D system on honeycomb lattice without spin orbital coupling, where $SU(2)$ spin rotational symmetry, space inversion and $C_3$ rotational symmetry are all preserved. With a beam of light, only a spin singlet pair of excitations at $\mathbf{k}$ and $-\mathbf{k}$ are excited. Inversion symmetry can also have selection rules in this case. The spectrum is zero or note depends on whether inversion is $I^{\prime2} = 1$ or $-1$.

To be specific, if $I^{\prime2} = 1$, we can choose a gauge such that under inversion $I^\prime$
\begin{equation}
\begin{aligned}
    I^\prime \gamma_{\mathbf{k}, \uparrow} I^{\prime-1} &= \gamma_{-\mathbf{k}, \uparrow} \\
    I^\prime \gamma_{\mathbf{k}, \uparrow} I^{\prime-1} &= \gamma_{-\mathbf{k}, \uparrow}
\end{aligned}
\end{equation}
This leads to the action of inversion on the excitation of singlet pair
\begin{equation}
    I^\prime (\gamma_{\mathbf{k}, \uparrow} \gamma_{-\mathbf{k}, \downarrow} - \gamma_{\mathbf{k}, \downarrow} \gamma_{-\mathbf{k}, \uparrow}) I ^{\prime-1} = \gamma_{-\mathbf{k}, \uparrow} \gamma_{\mathbf{k}, \downarrow} - \gamma_{-\mathbf{k}, \downarrow} \gamma_{\mathbf{k}, \uparrow} = \gamma_{\mathbf{k}, \uparrow} \gamma_{-\mathbf{k}, \downarrow} - \gamma_{\mathbf{k}, \downarrow} \gamma_{-\mathbf{k}, \uparrow}.
\end{equation}
Therefore this excitation is even under inversion and there will be no absorption spectrum because current is odd $I^\prime \hat{j} I^{\prime-1} = -\hat{j}$.

On the other hand, if $I^{\prime2} = -1$, under a certain gauge, action of inversion on fermions are
\begin{equation}
\begin{aligned}
    I^\prime \gamma_{\mathbf{k}, \uparrow} I^{\prime-1} &= \gamma_{-\mathbf{k}, \uparrow} \\
    I^\prime \gamma_{\mathbf{k}, \uparrow} I^{\prime-1} &= -\gamma_{-\mathbf{k}, \uparrow}
\end{aligned}
\end{equation}
we have one extra minus sign and excitation is odd under inversion. In this case, optical absorption is allowed.

With this general rule in mind, we write down a tight-binding model on honeycomb lattice explicitly. Total hamiltonian will still be in the form of Eq. \ref{eq.Hk} and \ref{eq.Hk_operator}. The only difference is operator $\Psi_k = (c_{1, {\mathbf{k}}, \uparrow}, c_{2, {\mathbf{k}}, \uparrow}, c_{1, -{\mathbf{k}}, \downarrow}^\dagger, c_{2, -{\mathbf{k}}, \downarrow}^\dagger)^T$. Here label $1, 2$ is sublattice number.

We consider only on-site and nearest neighbour terms. Due to the $SU(2)$ spin rotational symmetry, hopping terms between two sites can only be $t \hat{c}^\dagger_{i, \uparrow} \hat{c}_{j, \uparrow} + t \hat{c}^\dagger_{i, \downarrow} \hat{c}_{j, \downarrow} + h.c.$. It is even under link center inversion when $t$ is real, and odd when $t$ is imaginary. Pairing term can only be $\Delta \hat{c}_{i, \uparrow} \hat{c}_{j, \downarrow} - \Delta \hat{c}_{i, \downarrow} \hat{c}_{j, \uparrow} + h.c.$ and is even under permutation of $i$ and $j$.

Hopping term of Hamiltonian is invariant under inversion and $C_3$ rotational symmetry:
\begin{equation}
    \begin{aligned}
        I : &\quad h_0(- \mathbf{k}) = \sigma^x h_0(\mathbf{k}) \sigma^x\\
        C_3 : &\quad h_0((-k_2, k_1 - k_2)) = S^\dagger h_0((k_1, k_2)) S
    \end{aligned}
\end{equation}
where $(k_1, k_2)$ are coordinates of the two reciprocal lattice vectors, meaning $\mathbf{k} = k_1 \hat{k}_1 + k_2 \hat{k}_2$. and $S$ is matrix
\begin{equation}
    S = \begin{bmatrix}
        1 & \\
        & e^{-ik_2}
    \end{bmatrix}
\end{equation}
Symmetry allowed hopping terms are chemical potential and $C_3$ invariant real hopping between nearest neighbours. The general form of hopping term of Hamiltonian in momentum space is
\begin{equation}
    h_0(\mathbf{k}) =
    \begin{bmatrix}
        \mu & t(1 + e^{i k_1} + e^{i k_2})\\
        t(1 + e^{-i k_1} + e^{-i k_2}) & \mu
    \end{bmatrix}
\end{equation}
Here, $t, \mu$ are real numbers.

Next, current operator $j^z$ is zero and $j^{x,y}$ are of the following form:
\begin{equation}
    \begin{aligned}
        j_x (\mathbf{k}) &=
        \begin{bmatrix}
             & i \frac{t}{2} ( e^{i k_1} - e^ {i k_2})\\
             -i \frac{t}{2} ( e^{-i k_1} - e^ {-i k_2}) &
        \end{bmatrix}\\
        j_y (\mathbf{k}) &=
        \begin{bmatrix}
             & i \frac{\sqrt{3}t}{6} (-2 + e^{i k_1} + e^ {i k_2})\\
              -i \frac{\sqrt{3}t}{6} (-2 + e^{-i k_1} + e^ {-i k_2}) &
        \end{bmatrix}
    \end{aligned}
\end{equation}

To write down pairing term of Hamiltonian explicitly, we need to fix inversion symmetry fractionalization. Below we consider two different kinds of pairing with $I^2 = \pm 1$.\\

Case 1: Inversion symmetry is $I^\prime = I$, i.e. $I^{\prime2} = 1$. Paring terms are invariant under symmetry $I^\prime$ and $C_3$ are onsite and nearest neighbour pairings of the same strength. Pairing terms in Hamiltonian are
\begin{equation}
    \Delta(\mathbf{k}) =
    \begin{bmatrix}
        \Delta & \Delta_1 ( 1 + e^{i k_1} + e ^ {i k_2}) \\
        \Delta_1 ^ * ( 1 + e^{-i k_1} + e ^ {-i k_2})  & \Delta
    \end{bmatrix}
\end{equation}
Parameters are $\mu = 0.5t, \ \Delta = \Delta_1 = 0.5t$.

Case 2: Inversion symmetry is $I^\prime = I \exp{(i \pi \hat{F} / 2 )}$, i.e. $I^{\prime 2} = -1$. Paring term in Hamiltonian is
\begin{equation}
    \Delta(\mathbf{k}) = \Delta_2 \sigma^z
\end{equation}
Parameters are $\mu = t, \  \Delta_2 = 4t$.
\begin{figure}[h]
    \centering
    \includegraphics[width = 0.48 \columnwidth]{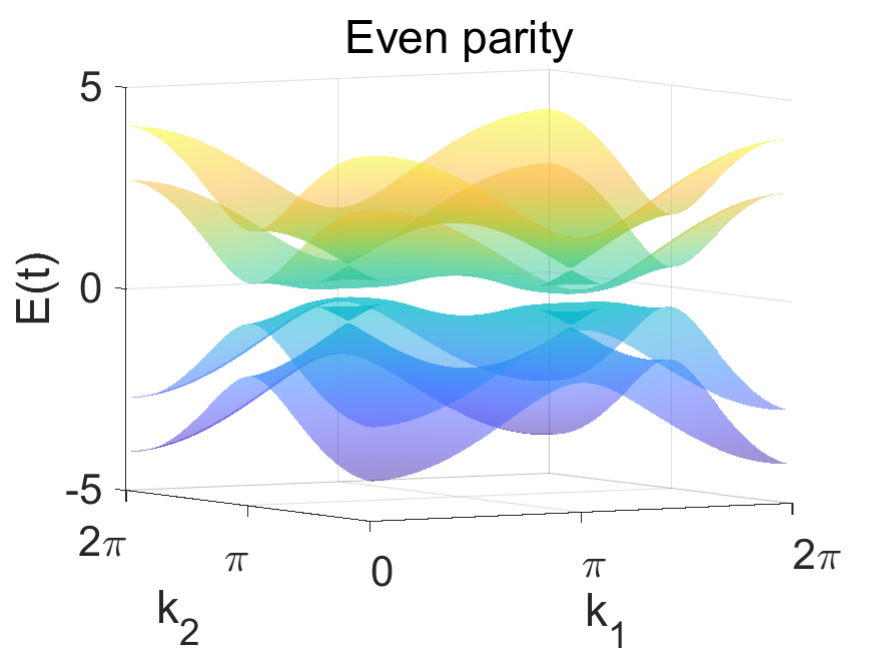}
    \includegraphics[width = 0.48 \columnwidth]{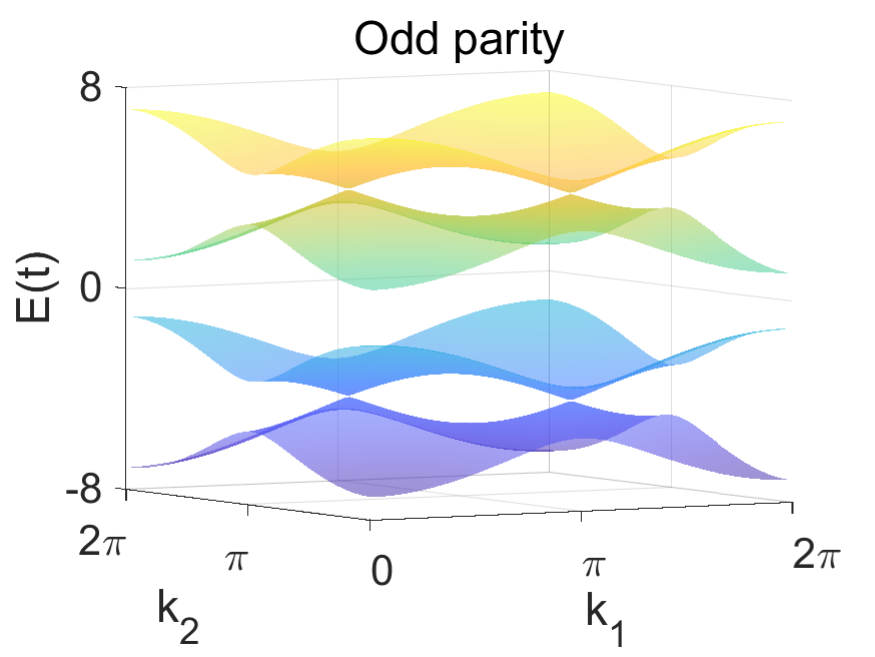}
    \caption{Electron bands of two cases. Even parity case has a small gap. Energy gap of first band is $\Delta = 0.22 t$ and second band gap is $\Delta^\prime = 0.75 t$. Odd parity case is gapless with first band gap $\Delta = 1.1 t$ and gap of second band $\Delta^\prime = 4.2 t$.}.
    \label{fig:hex_band}
\end{figure}

\begin{figure}[h]
    \centering
    \includegraphics[width = 0.48 \columnwidth]{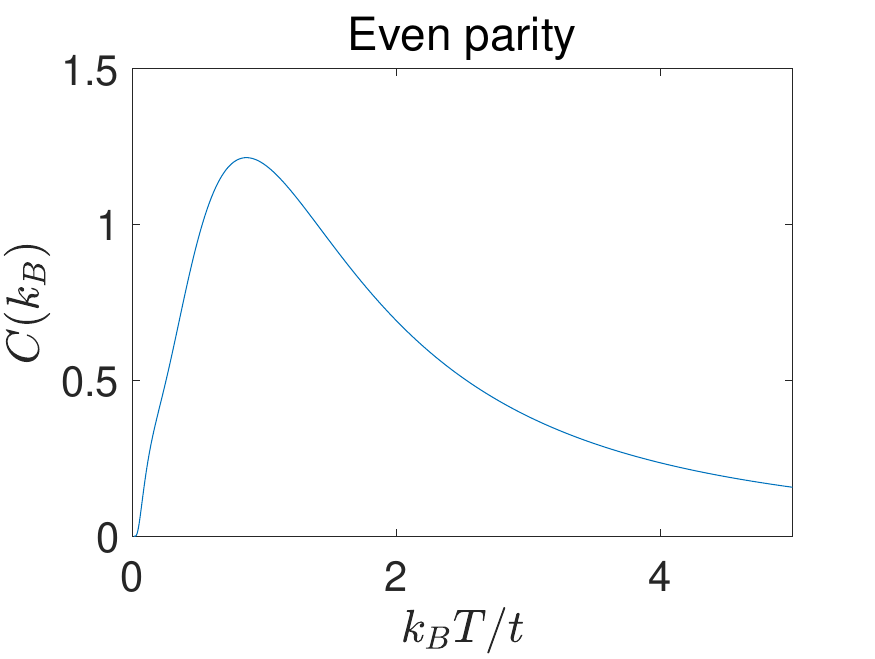}
    \includegraphics[width = 0.48 \columnwidth]{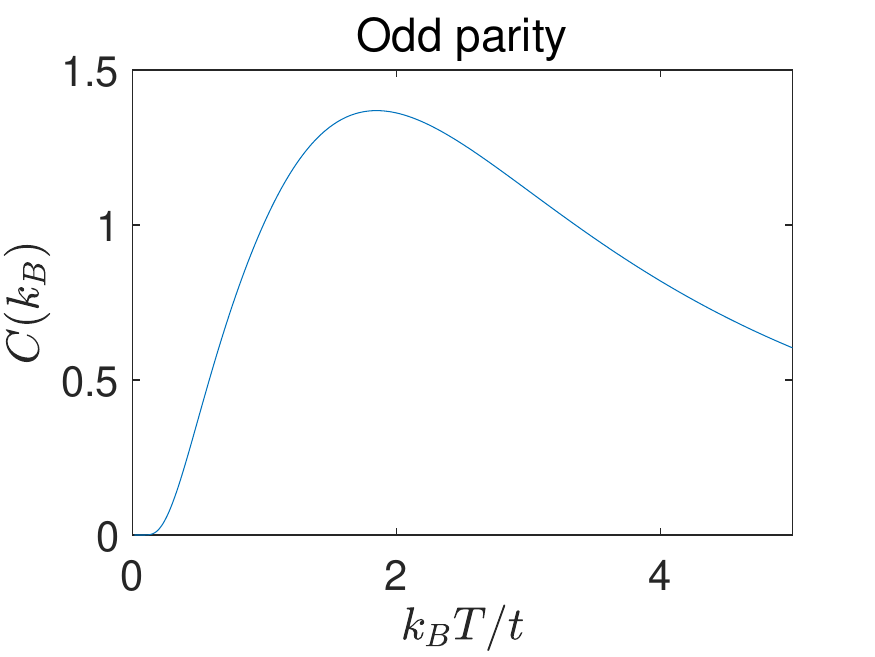}
    \caption{specific heat versus temperature $T$. Both the even-parity and odd-parity cases are gapped and specific heat are exponential function of $1/T$ at low temperature. Energy gap is $\Delta = 0.22 t$ for even-parity case and $\Delta = 0.56 t$ for odd-parity case.}
    \label{fig:hex_heat_capacity}
\end{figure}

Now we calculate observable quantities for this mode. The calculation is the same as the case without $SU(2)$ symmetry. 

We show electron bands of two cases in Fig. \ref{fig:hex_band} and specific heat in Fig.\ref{fig:hex_heat_capacity}. Both cases are gapped and specific heat shows $C \approx \exp{(\Delta/T)}$ at low temperature. 

\begin{figure}[h]
    \centering
    \includegraphics[width = 0.48 \columnwidth]{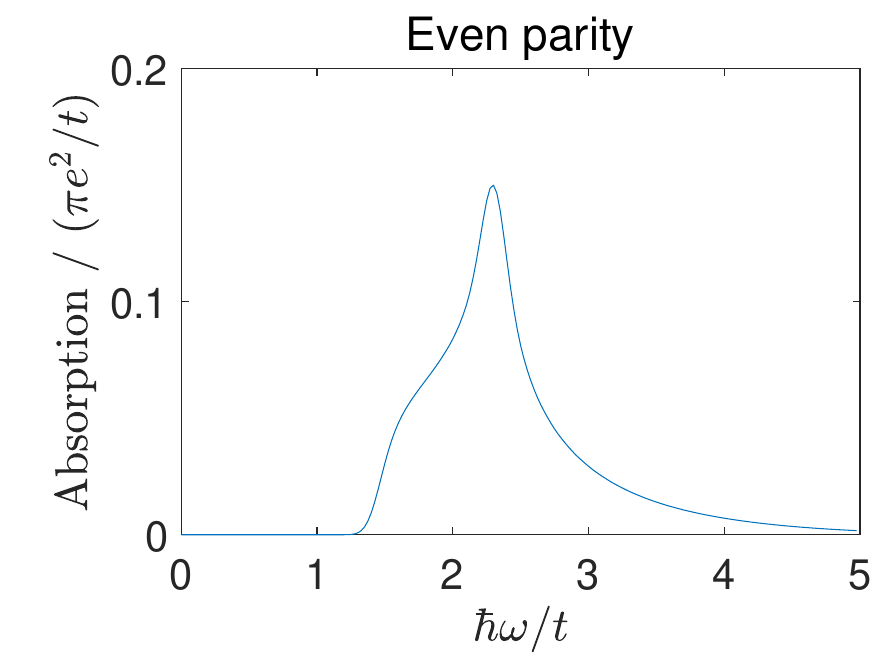}
    \includegraphics[width = 0.48 \columnwidth]{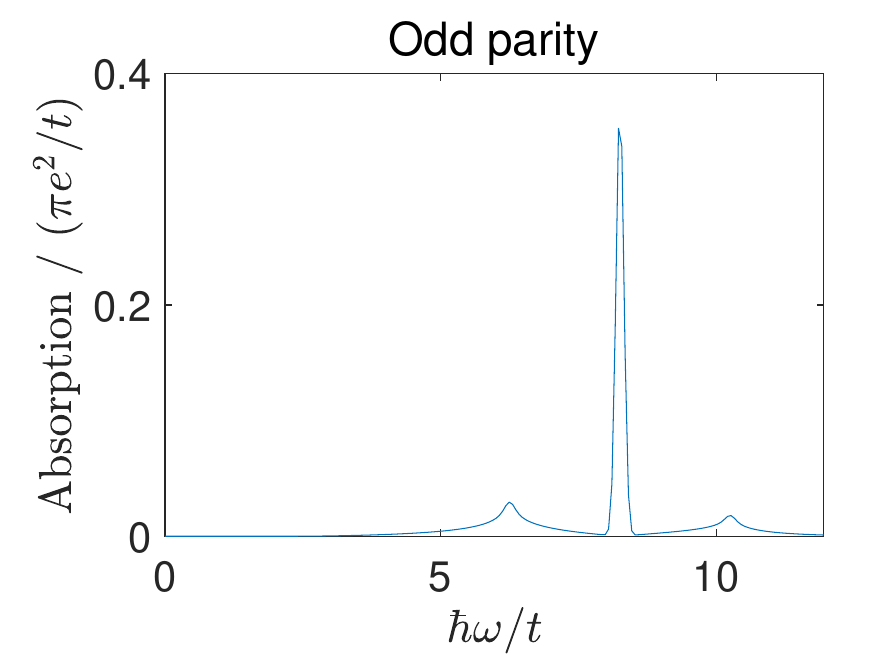}
    \caption{Absorption spectrum of models with $I^{\prime2} = 1$ and polarization in $x$ direction (left). In comparison, we show the spectrum of model with $I^{\prime2} = -1$ and polarization in $x$ direction (right).}
    \label{fig:spectrum_su2}
\end{figure}

Absorption spectrum can be calculated based on Eq. \ref{eq.spectrum}. We use light polarized in $x$ direction. Absorption spectrum of these two cases is shown in Fig. \ref{fig:spectrum_su2}. On the left are models with $I^{\prime2} = 1$, and on the right are $I^{\prime2} = -1$.  It is clear the spectrum satisfies the selection rule for even-parity case. For even-parity case, spectrum is zero from energy gap $2\Delta = 0.44 t$ to $\Delta + \Delta^\prime = 0.97 t$. For odd-parity case, the spectrum is not zero above energy gap $2\Delta = 2.2 t$.

\begin{figure}[h]
    \centering
    \includegraphics[width = 0.48 \columnwidth]{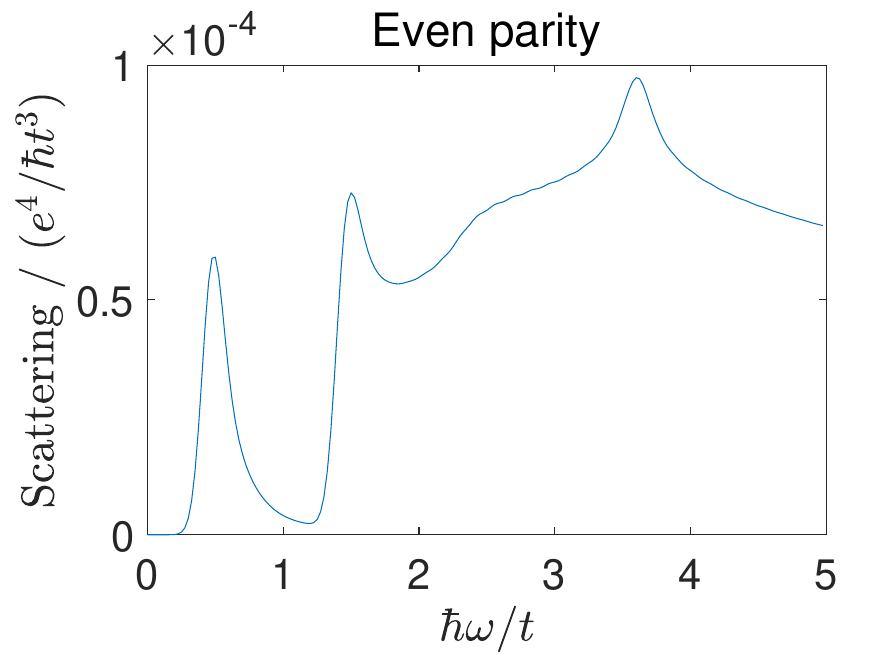}
    \includegraphics[width = 0.48 \columnwidth]{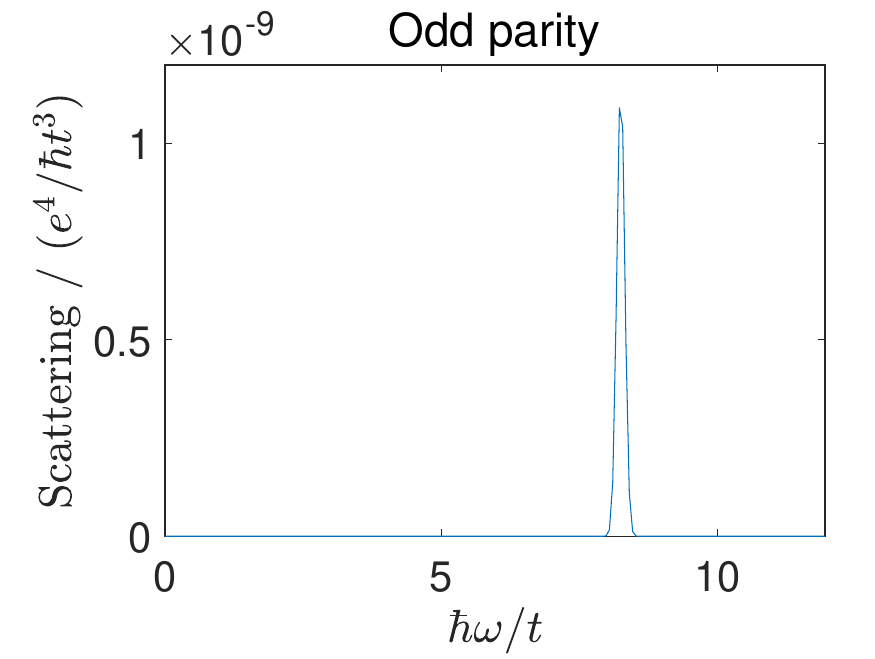}
    \caption{Differential scattering cross section of both incident and scattered light polarized in $x$ direction. Figures are corresponding to $I^2 = 1$ (left) and $I^2 = -1$ (right).}
    \label{fig:raman_su2}
\end{figure}

Raman spectrum is shown in Fig. \ref{fig:raman_su2}. Odd-parity case shows selection rule. For even-parity case, scattering is not zero above energy gap $2\Delta$. But for odd-parity case, scattering is zero between energy gap $2\Delta$ and $\Delta + \Delta^\prime$. 

\section{Discussion of selection rules for general point group symmetries}
In this section we systematically study possible selection rules in the case of a general point group. Our discussion includes the case of a unitary point group $G_{total}=G$ and the case of a magnetic point group that can be written as $G_{total}=G+AG$, where $G$ is the unitary subgroup and $AG$ is the coset corresponding to an anti-unitary symmetry A (including but not restricted to time-reversal symmetry). 

In the same spirit as the case of inversion symmetry, we want to ask whether the strength of spectrum contributed by pair excitation from the lowest bands is constrained to vanish by symmetry. Specifically, we want to study for general momentum $\mathbf{k}$ whether the following matrix element
\bea\label{appendeq:matrixelement}
    \expval{ 0 | \hat{O} ~ \gamma^\dagger_{\mathbf{k},0}\gamma^\dagger_{-\mathbf{k},0}| 0}
\eea
is zero. Here $\hat{O}$ represents current operator $\hat{j}$ for absorption spectrum and Raman operator $R$ for Raman spectrum. To study effect of symmetry on this, we study representation of pair excitation states 
\bea
\dket{f_\mathbf{k}}=\gamma^\dagger_{\mathbf{k},0}\gamma^\dagger_{-\mathbf{k},0}\dket{0}
\eea
We decompose representation $\{ \dket{f_{g(\mathbf{k})}}, g \in G_{total} \}$ into irreducible representations and check whether it can form a trivial representation after combining with operator $\hat{O}$. If it cannot, the matrix element transforms non-trivially under the symmetry group and is constrained to be zero-hence we have the selection rule. Otherwise, the spectrum does not vanish. In the special case of $\{ \dket{f_{\mathbf{k}}} \}$ having all the representations of group $G_{total}$, there is no selection rule since the result does not depend on representations of operator $\hat{O}$. 

Let's consider a general momentum $\mathbf{k}$. The only symmetries that does not change the $\mathbf{k}, -\mathbf{k}$ pair are inversion $I$, time-reversal symmetry $T$, and their product $TI$. In the following, we divide the problem into several situations and discuss in detail.

Case 1: Unitary point group G with inversion $I \notin G$. In this case, $\{ \dket{f_{\mathbf{k}}} \}$ forms unitary representations of group $G$. Since inversion is not in $G$, all $g(\mathbf{k})$ are different and $\{ \dket{f_{\mathbf{k}}} \}$ forms the $|G|$ dimensional fundamental representation of $G$. Since fundamental representation contains all irreducible representations, there is no selection rule in this case.

Case 2: Unitary point group G with inversion $I \in G$. In this case, every group element other than $I$ will change momentum. So $\{ \dket{f_{\mathbf{k}}} \}$ forms fundamental representation of $G / Z_2^I$ ($Z_2^I$ means $Z_2$ group of inversion). As representation of $G$, depending on $I = + 1$ or $I = -1$, $\{ \dket{f_{\mathbf{k}}} \}$ contains half of irreducible representations and misses the other half. This gives us the selection rule of inversion symmetry, as discussed before. There is no other selection rules.

Below we discuss the case of magnetic point groups with anti-unitary symmetries, where $\{ \dket{f_{\mathbf{k}}} \}$ forms co-representation of the symmetry group \cite{Bradley1968}. One property we will use is the co-representations of $G + AG$ has a one to one correspondence with the representations of the unitary subgroup $G$. To see which co-representation is contained, we can simply forget about anti-unitary part and see which representation of $G$ is contained.

Case 3: Magnetic point group $G_{total} = G + AG$ with inversion $I \notin G$. In this case, $\{ \dket{f_{\mathbf{k}}} \}$ forms fundamental representation of $G$ as in case 1. Therefore, it contains every co-representation of $G_{total}$ and there is no selection rule.

Case 4: Magnetic point group $G_{total} = G + AG$ with inversion $I \in G$ and time-reversal $T \notin AG$. Following the same logic, this case is the same as case 2. $\{ \dket{f_{\mathbf{k}}} \}$ contains half of irreducible co-representations of $G_{total}$ depending on $I = + 1$ or $I = -1$. This gives us selection rule of inversion symmetry.

Case 5: Magnetic point group $G_{total} = G + AG$ with inversion $I \in G$ and time-reversal $T \in AG$. This case is special because $IT$ guarantees a two-fold band degeneracy. We have states 
\bea
\dket{f_{\mathbf{k}, \alpha, \beta}}=\gamma^\dagger_{\mathbf{k},0, \alpha}\gamma^\dagger_{-\mathbf{k},0, \beta}\dket{0}
\eea
where $\alpha, \beta = 1, 2$ is label of degenerate bands. We pick up states with $\alpha = 1, \beta = 2$ and realize $\dket{f_{\mathbf{k}, 1, 2}}$ and $\dket{f_{-\mathbf{k}, 1, 2}}$ are two different states. So,  $\{ \dket{f_{\mathbf{k}, 1, 2}} \}$ forms fundamental representation of $G$. Since $\{ \dket{f_{\mathbf{k}, 1, 2}} \}$ already has all irreducible representations, the matrix element \eqref{appendeq:matrixelement} is non-zero and there is no selection rule in this case.

In conclusion, selection rules only arise for inversion symmetry (and $C_{2,z}$ symmetry in 2d, which acts like an inversion symmetry). No other crystal symmetries or magnetic crystal symmetries lead to nontrivial selection rules in optical and Raman spectroscopy.

\bibliographystyle{apsrev}
\bibliography{SC_selection_rules}

\begin{thebibliography}{44}
\expandafter\ifx\csname natexlab\endcsname\relax\def\natexlab#1{#1}\fi
\expandafter\ifx\csname bibnamefont\endcsname\relax
  \def\bibnamefont#1{#1}\fi
\expandafter\ifx\csname bibfnamefont\endcsname\relax
  \def\bibfnamefont#1{#1}\fi
\expandafter\ifx\csname citenamefont\endcsname\relax
  \def\citenamefont#1{#1}\fi
\expandafter\ifx\csname url\endcsname\relax
  \def\url#1{\texttt{#1}}\fi
\expandafter\ifx\csname urlprefix\endcsname\relax\def\urlprefix{URL }\fi
\providecommand{\bibinfo}[2]{#2}
\providecommand{\eprint}[2][]{\url{#2}}

\bibitem[{\citenamefont{Degen et~al.}(2017)\citenamefont{Degen, Reinhard, and
  Cappellaro}}]{Degen2017}
\bibinfo{author}{\bibfnamefont{C.~L.} \bibnamefont{Degen}},
  \bibinfo{author}{\bibfnamefont{F.}~\bibnamefont{Reinhard}}, \bibnamefont{and}
  \bibinfo{author}{\bibfnamefont{P.}~\bibnamefont{Cappellaro}},
  \bibinfo{journal}{Rev. Mod. Phys.} \textbf{\bibinfo{volume}{89}},
  \bibinfo{pages}{035002} (\bibinfo{year}{2017}),
  \urlprefix\url{https://link.aps.org/doi/10.1103/RevModPhys.89.035002}.

\bibitem[{\citenamefont{Kjaergaard et~al.}(2023)\citenamefont{Kjaergaard,
  Schwartz, Braumï¿½ller, Krantz, Wang, Gustavsson, and
  Oliver}}]{Kjaergaard2023}
\bibinfo{author}{\bibfnamefont{M.}~\bibnamefont{Kjaergaard}},
  \bibinfo{author}{\bibfnamefont{M.~E.} \bibnamefont{Schwartz}},
  \bibinfo{author}{\bibfnamefont{J.}~\bibnamefont{Braumï¿½ller}},
  \bibinfo{author}{\bibfnamefont{P.}~\bibnamefont{Krantz}},
  \bibinfo{author}{\bibfnamefont{J.~I.-J.} \bibnamefont{Wang}},
  \bibinfo{author}{\bibfnamefont{S.}~\bibnamefont{Gustavsson}},
  \bibnamefont{and} \bibinfo{author}{\bibfnamefont{W.~D.}
  \bibnamefont{Oliver}}, \bibinfo{journal}{Annu. Rev. Condens. Matter Phys.}
  \textbf{\bibinfo{volume}{11}}, \bibinfo{pages}{369} (\bibinfo{year}{2023}),
  ISSN \bibinfo{issn}{1947-5454},
  \urlprefix\url{https://doi.org/10.1146/annurev-conmatphys-031119-050605}.

\bibitem[{\citenamefont{MacManus-Driscoll and
  Wimbush}(2021)}]{MacManus-Driscoll2021}
\bibinfo{author}{\bibfnamefont{J.~L.} \bibnamefont{MacManus-Driscoll}}
  \bibnamefont{and} \bibinfo{author}{\bibfnamefont{S.~C.}
  \bibnamefont{Wimbush}}, \bibinfo{journal}{Nature Reviews Materials}
  \textbf{\bibinfo{volume}{6}}, \bibinfo{pages}{587} (\bibinfo{year}{2021}),
  ISSN \bibinfo{issn}{2058-8437},
  \urlprefix\url{https://doi.org/10.1038/s41578-021-00290-3}.

\bibitem[{\citenamefont{Annett}(1990)}]{Annett1990}
\bibinfo{author}{\bibfnamefont{J.~F.} \bibnamefont{Annett}},
  \bibinfo{journal}{Advances in Physics} \textbf{\bibinfo{volume}{39}},
  \bibinfo{pages}{83} (\bibinfo{year}{1990}), ISSN \bibinfo{issn}{0001-8732},
  \urlprefix\url{http://www.tandfonline.com/doi/abs/10.1080/00018739000101481}.

\bibitem[{\citenamefont{Sigrist and Ueda}(1991)}]{Sigrist1991}
\bibinfo{author}{\bibfnamefont{M.}~\bibnamefont{Sigrist}} \bibnamefont{and}
  \bibinfo{author}{\bibfnamefont{K.}~\bibnamefont{Ueda}},
  \bibinfo{journal}{Rev. Mod. Phys.} \textbf{\bibinfo{volume}{63}},
  \bibinfo{pages}{239} (\bibinfo{year}{1991}),
  \urlprefix\url{http://link.aps.org/doi/10.1103/RevModPhys.63.239}.

\bibitem[{\citenamefont{Tsuei and Kirtley}(2000)}]{Tsuei2000}
\bibinfo{author}{\bibfnamefont{C.~C.} \bibnamefont{Tsuei}} \bibnamefont{and}
  \bibinfo{author}{\bibfnamefont{J.~R.} \bibnamefont{Kirtley}},
  \bibinfo{journal}{Rev. Mod. Phys.} \textbf{\bibinfo{volume}{72}},
  \bibinfo{pages}{969} (\bibinfo{year}{2000}),
  \urlprefix\url{http://link.aps.org/doi/10.1103/RevModPhys.72.969}.

\bibitem[{\citenamefont{Damascelli et~al.}(2003)\citenamefont{Damascelli,
  Hussain, and Shen}}]{Damascelli2003}
\bibinfo{author}{\bibfnamefont{A.}~\bibnamefont{Damascelli}},
  \bibinfo{author}{\bibfnamefont{Z.}~\bibnamefont{Hussain}}, \bibnamefont{and}
  \bibinfo{author}{\bibfnamefont{Z.-X.} \bibnamefont{Shen}},
  \bibinfo{journal}{Rev. Mod. Phys.} \textbf{\bibinfo{volume}{75}},
  \bibinfo{pages}{473} (\bibinfo{year}{2003}),
  \urlprefix\url{https://link.aps.org/doi/10.1103/RevModPhys.75.473}.

\bibitem[{\citenamefont{Lee et~al.}(2006)\citenamefont{Lee, Nagaosa, and
  Wen}}]{Lee2006}
\bibinfo{author}{\bibfnamefont{P.~A.} \bibnamefont{Lee}},
  \bibinfo{author}{\bibfnamefont{N.}~\bibnamefont{Nagaosa}}, \bibnamefont{and}
  \bibinfo{author}{\bibfnamefont{X.-G.} \bibnamefont{Wen}},
  \bibinfo{journal}{Rev. Mod. Phys.} \textbf{\bibinfo{volume}{78}},
  \bibinfo{pages}{17} (\bibinfo{year}{2006}),
  \urlprefix\url{http://journals.aps.org/rmp/abstract/10.1103/RevModPhys.78.17}.

\bibitem[{\citenamefont{Taillefer}(2010)}]{Taillefer2010}
\bibinfo{author}{\bibfnamefont{L.}~\bibnamefont{Taillefer}},
  \bibinfo{journal}{Annu. Rev. Condens. Matter Phys.}
  \textbf{\bibinfo{volume}{1}}, \bibinfo{pages}{51} (\bibinfo{year}{2010}),
  ISSN \bibinfo{issn}{1947-5454},
  \urlprefix\url{https://doi.org/10.1146/annurev-conmatphys-070909-104117}.

\bibitem[{\citenamefont{Armitage et~al.}(2010)\citenamefont{Armitage, Fournier,
  and Greene}}]{Armitage2010}
\bibinfo{author}{\bibfnamefont{N.~P.} \bibnamefont{Armitage}},
  \bibinfo{author}{\bibfnamefont{P.}~\bibnamefont{Fournier}}, \bibnamefont{and}
  \bibinfo{author}{\bibfnamefont{R.~L.} \bibnamefont{Greene}},
  \bibinfo{journal}{Rev. Mod. Phys.} \textbf{\bibinfo{volume}{82}},
  \bibinfo{pages}{2421} (\bibinfo{year}{2010}),
  \urlprefix\url{http://link.aps.org/doi/10.1103/RevModPhys.82.2421}.

\bibitem[{\citenamefont{Ando and Fu}(2015)}]{Ando2015}
\bibinfo{author}{\bibfnamefont{Y.}~\bibnamefont{Ando}} \bibnamefont{and}
  \bibinfo{author}{\bibfnamefont{L.}~\bibnamefont{Fu}}, \bibinfo{journal}{Annu.
  Rev. Condens. Matter Phys.} \textbf{\bibinfo{volume}{6}},
  \bibinfo{pages}{361} (\bibinfo{year}{2015}), ISSN \bibinfo{issn}{1947-5454},
  \urlprefix\url{http://dx.doi.org/10.1146/annurev-conmatphys-031214-014501}.

\bibitem[{\citenamefont{Sato and Ando}(2017)}]{Sato2017}
\bibinfo{author}{\bibfnamefont{M.}~\bibnamefont{Sato}} \bibnamefont{and}
  \bibinfo{author}{\bibfnamefont{Y.}~\bibnamefont{Ando}},
  \bibinfo{journal}{Reports on Progress in Physics}
  \textbf{\bibinfo{volume}{80}}, \bibinfo{pages}{076501}
  (\bibinfo{year}{2017}), ISSN \bibinfo{issn}{0034-4885},
  \urlprefix\url{http://stacks.iop.org/0034-4885/80/i=7/a=076501}.

\bibitem[{\citenamefont{Nayak et~al.}(2008)\citenamefont{Nayak, Simon, Stern,
  Freedman, and Das~Sarma}}]{Nayak2008}
\bibinfo{author}{\bibfnamefont{C.}~\bibnamefont{Nayak}},
  \bibinfo{author}{\bibfnamefont{S.~H.} \bibnamefont{Simon}},
  \bibinfo{author}{\bibfnamefont{A.}~\bibnamefont{Stern}},
  \bibinfo{author}{\bibfnamefont{M.}~\bibnamefont{Freedman}}, \bibnamefont{and}
  \bibinfo{author}{\bibfnamefont{S.}~\bibnamefont{Das~Sarma}},
  \bibinfo{journal}{Rev. Mod. Phys.} \textbf{\bibinfo{volume}{80}},
  \bibinfo{pages}{1083} (\bibinfo{year}{2008}),
  \urlprefix\url{http://link.aps.org/doi/10.1103/RevModPhys.80.1083}.

\bibitem[{\citenamefont{Alicea}(2012)}]{Alicea2012}
\bibinfo{author}{\bibfnamefont{J.}~\bibnamefont{Alicea}},
  \bibinfo{journal}{Reports on Progress in Physics}
  \textbf{\bibinfo{volume}{75}}, \bibinfo{pages}{076501}
  (\bibinfo{year}{2012}),
  \urlprefix\url{http://stacks.iop.org/0034-4885/75/i=7/a=076501}.

\bibitem[{\citenamefont{Beenakker}(2013)}]{Beenakker2013}
\bibinfo{author}{\bibfnamefont{C.}~\bibnamefont{Beenakker}},
  \bibinfo{journal}{Annu. Rev. Condens. Matter Phys.}
  \textbf{\bibinfo{volume}{4}}, \bibinfo{pages}{113} (\bibinfo{year}{2013}),
  ISSN \bibinfo{issn}{1947-5454},
  \urlprefix\url{http://dx.doi.org/10.1146/annurev-conmatphys-030212-184337}.

\bibitem[{\citenamefont{Lutchyn et~al.}(2018)\citenamefont{Lutchyn, Bakkers,
  Kouwenhoven, Krogstrup, Marcus, and Oreg}}]{Lutchyn2018}
\bibinfo{author}{\bibfnamefont{R.~M.} \bibnamefont{Lutchyn}},
  \bibinfo{author}{\bibfnamefont{E.~P. A.~M.} \bibnamefont{Bakkers}},
  \bibinfo{author}{\bibfnamefont{L.~P.} \bibnamefont{Kouwenhoven}},
  \bibinfo{author}{\bibfnamefont{P.}~\bibnamefont{Krogstrup}},
  \bibinfo{author}{\bibfnamefont{C.~M.} \bibnamefont{Marcus}},
  \bibnamefont{and} \bibinfo{author}{\bibfnamefont{Y.}~\bibnamefont{Oreg}},
  \bibinfo{journal}{Nature Reviews Materials} \textbf{\bibinfo{volume}{3}},
  \bibinfo{pages}{52} (\bibinfo{year}{2018}), ISSN \bibinfo{issn}{2058-8437},
  \urlprefix\url{https://doi.org/10.1038/s41578-018-0003-1}.

\bibitem[{\citenamefont{Van~Harlingen}(1995)}]{VanHarlingen1995}
\bibinfo{author}{\bibfnamefont{D.~J.} \bibnamefont{Van~Harlingen}},
  \bibinfo{journal}{Rev. Mod. Phys.} \textbf{\bibinfo{volume}{67}},
  \bibinfo{pages}{515} (\bibinfo{year}{1995}),
  \urlprefix\url{http://link.aps.org/doi/10.1103/RevModPhys.67.515}.

\bibitem[{\citenamefont{Scalapino}(1995)}]{Scalapino1995}
\bibinfo{author}{\bibfnamefont{D.~J.} \bibnamefont{Scalapino}},
  \bibinfo{journal}{Physics Reports} \textbf{\bibinfo{volume}{250}},
  \bibinfo{pages}{329} (\bibinfo{year}{1995}), ISSN \bibinfo{issn}{0370-1573},
  \urlprefix\url{https://www.sciencedirect.com/science/article/pii/037015739400086I}.

\bibitem[{\citenamefont{Read and Green}(2000)}]{Read2000}
\bibinfo{author}{\bibfnamefont{N.}~\bibnamefont{Read}} \bibnamefont{and}
  \bibinfo{author}{\bibfnamefont{D.}~\bibnamefont{Green}},
  \bibinfo{journal}{Phys. Rev. B} \textbf{\bibinfo{volume}{61}},
  \bibinfo{pages}{10267} (\bibinfo{year}{2000}),
  \urlprefix\url{http://link.aps.org/doi/10.1103/PhysRevB.61.10267}.

\bibitem[{\citenamefont{{Yang} et~al.}(2023)\citenamefont{{Yang}, {Lu},
  {Biswas}, {Randeria}, and {Lu}}}]{Yang2024}
\bibinfo{author}{\bibfnamefont{X.}~\bibnamefont{{Yang}}},
  \bibinfo{author}{\bibfnamefont{S.}~\bibnamefont{{Lu}}},
  \bibinfo{author}{\bibfnamefont{S.}~\bibnamefont{{Biswas}}},
  \bibinfo{author}{\bibfnamefont{M.}~\bibnamefont{{Randeria}}},
  \bibnamefont{and} \bibinfo{author}{\bibfnamefont{Y.-M.} \bibnamefont{{Lu}}},
  \bibinfo{journal}{arXiv e-prints} \bibinfo{eid}{arXiv:2401.00321}
  (\bibinfo{year}{2023}), \eprint{2401.00321}.

\bibitem[{\citenamefont{Cao et~al.}(2018)\citenamefont{Cao, Fatemi, Fang,
  Watanabe, Taniguchi, Kaxiras, and Jarillo-Herrero}}]{Cao2018}
\bibinfo{author}{\bibfnamefont{Y.}~\bibnamefont{Cao}},
  \bibinfo{author}{\bibfnamefont{V.}~\bibnamefont{Fatemi}},
  \bibinfo{author}{\bibfnamefont{S.}~\bibnamefont{Fang}},
  \bibinfo{author}{\bibfnamefont{K.}~\bibnamefont{Watanabe}},
  \bibinfo{author}{\bibfnamefont{T.}~\bibnamefont{Taniguchi}},
  \bibinfo{author}{\bibfnamefont{E.}~\bibnamefont{Kaxiras}}, \bibnamefont{and}
  \bibinfo{author}{\bibfnamefont{P.}~\bibnamefont{Jarillo-Herrero}},
  \bibinfo{journal}{Nature} \textbf{\bibinfo{volume}{556}}, \bibinfo{pages}{43}
  (\bibinfo{year}{2018}), ISSN \bibinfo{issn}{1476-4687},
  \urlprefix\url{https://doi.org/10.1038/nature26160}.

\bibitem[{\citenamefont{Shiozaki}(2022)}]{Shiozaki2022}
\bibinfo{author}{\bibfnamefont{K.}~\bibnamefont{Shiozaki}},
  \bibinfo{journal}{Progress of Theoretical and Experimental Physics}
  \textbf{\bibinfo{volume}{2022}}, \bibinfo{pages}{04A104}
  (\bibinfo{year}{2022}), ISSN \bibinfo{issn}{2050-3911},
  \eprint{https://academic.oup.com/ptep/article-pdf/2022/4/04A104/43471665/ptac026.pdf},
  \urlprefix\url{https://doi.org/10.1093/ptep/ptep026}.

\bibitem[{\citenamefont{Ahn and Nagaosa}(2021)}]{Ahn2021}
\bibinfo{author}{\bibfnamefont{J.}~\bibnamefont{Ahn}} \bibnamefont{and}
  \bibinfo{author}{\bibfnamefont{N.}~\bibnamefont{Nagaosa}},
  \bibinfo{journal}{Nature Communications} \textbf{\bibinfo{volume}{12}},
  \bibinfo{pages}{1617} (\bibinfo{year}{2021}), ISSN \bibinfo{issn}{2041-1723},
  \urlprefix\url{https://doi.org/10.1038/s41467-021-21905-x}.

\bibitem[{\citenamefont{Devereaux and Hackl}(2007)}]{Devereaux2007}
\bibinfo{author}{\bibfnamefont{T.~P.} \bibnamefont{Devereaux}}
  \bibnamefont{and} \bibinfo{author}{\bibfnamefont{R.}~\bibnamefont{Hackl}},
  \bibinfo{journal}{Rev. Mod. Phys.} \textbf{\bibinfo{volume}{79}},
  \bibinfo{pages}{175} (\bibinfo{year}{2007}),
  \urlprefix\url{https://link.aps.org/doi/10.1103/RevModPhys.79.175}.

\bibitem[{\citenamefont{Schwarz et~al.}(2020)\citenamefont{Schwarz, Fauseweh,
  Tsuji, Cheng, Bittner, Krull, Berciu, Uhrig, Schnyder, Kaiser
  et~al.}}]{Schwarz2020}
\bibinfo{author}{\bibfnamefont{L.}~\bibnamefont{Schwarz}},
  \bibinfo{author}{\bibfnamefont{B.}~\bibnamefont{Fauseweh}},
  \bibinfo{author}{\bibfnamefont{N.}~\bibnamefont{Tsuji}},
  \bibinfo{author}{\bibfnamefont{N.}~\bibnamefont{Cheng}},
  \bibinfo{author}{\bibfnamefont{N.}~\bibnamefont{Bittner}},
  \bibinfo{author}{\bibfnamefont{H.}~\bibnamefont{Krull}},
  \bibinfo{author}{\bibfnamefont{M.}~\bibnamefont{Berciu}},
  \bibinfo{author}{\bibfnamefont{G.~S.} \bibnamefont{Uhrig}},
  \bibinfo{author}{\bibfnamefont{A.~P.} \bibnamefont{Schnyder}},
  \bibinfo{author}{\bibfnamefont{S.}~\bibnamefont{Kaiser}},
  \bibnamefont{et~al.}, \bibinfo{journal}{Nature Communications}
  \textbf{\bibinfo{volume}{11}}, \bibinfo{pages}{287} (\bibinfo{year}{2020}),
  ISSN \bibinfo{issn}{2041-1723},
  \urlprefix\url{https://doi.org/10.1038/s41467-019-13763-5}.

\bibitem[{\citenamefont{Shimano and Tsuji}(2023)}]{Shimano2023}
\bibinfo{author}{\bibfnamefont{R.}~\bibnamefont{Shimano}} \bibnamefont{and}
  \bibinfo{author}{\bibfnamefont{N.}~\bibnamefont{Tsuji}},
  \bibinfo{journal}{Annu. Rev. Condens. Matter Phys.}
  \textbf{\bibinfo{volume}{11}}, \bibinfo{pages}{103} (\bibinfo{year}{2023}),
  ISSN \bibinfo{issn}{1947-5454},
  \urlprefix\url{https://doi.org/10.1146/annurev-conmatphys-031119-050813}.

\bibitem[{\citenamefont{Mahan}(2000)}]{Mahan2000B}
\bibinfo{author}{\bibfnamefont{G.~D.} \bibnamefont{Mahan}},
  \emph{\bibinfo{title}{Superconductivity}} (\bibinfo{publisher}{Springer US},
  \bibinfo{address}{Boston, MA}, \bibinfo{year}{2000}), pp.
  \bibinfo{pages}{627--675}, ISBN \bibinfo{isbn}{978-1-4757-5714-9},
  \urlprefix\url{https://doi.org/10.1007/978-1-4757-5714-9_10}.

\bibitem[{\citenamefont{Yang et~al.}(2011)\citenamefont{Yang, Lu, and
  Ran}}]{Yang2011}
\bibinfo{author}{\bibfnamefont{K.-Y.} \bibnamefont{Yang}},
  \bibinfo{author}{\bibfnamefont{Y.-M.} \bibnamefont{Lu}}, \bibnamefont{and}
  \bibinfo{author}{\bibfnamefont{Y.}~\bibnamefont{Ran}},
  \bibinfo{journal}{Phys. Rev. B} \textbf{\bibinfo{volume}{84}},
  \bibinfo{pages}{075129} (\bibinfo{year}{2011}),
  \urlprefix\url{http://link.aps.org/doi/10.1103/PhysRevB.84.075129}.

\bibitem[{\citenamefont{Cho et~al.}(2012)\citenamefont{Cho, Bardarson, Lu, and
  Moore}}]{Cho2012}
\bibinfo{author}{\bibfnamefont{G.~Y.} \bibnamefont{Cho}},
  \bibinfo{author}{\bibfnamefont{J.~H.} \bibnamefont{Bardarson}},
  \bibinfo{author}{\bibfnamefont{Y.-M.} \bibnamefont{Lu}}, \bibnamefont{and}
  \bibinfo{author}{\bibfnamefont{J.~E.} \bibnamefont{Moore}},
  \bibinfo{journal}{Phys. Rev. B} \textbf{\bibinfo{volume}{86}},
  \bibinfo{pages}{214514} (\bibinfo{year}{2012}),
  \urlprefix\url{https://link.aps.org/doi/10.1103/PhysRevB.86.214514}.

\bibitem[{Sup(2023)}]{Supp}
\emph{\bibinfo{title}{For details see supplemental materials.}}
  (\bibinfo{year}{2023}).

\bibitem[{\citenamefont{Li and Haldane}(2018)}]{Li2018a}
\bibinfo{author}{\bibfnamefont{Y.}~\bibnamefont{Li}} \bibnamefont{and}
  \bibinfo{author}{\bibfnamefont{F.~D.~M.} \bibnamefont{Haldane}},
  \bibinfo{journal}{Phys. Rev. Lett.} \textbf{\bibinfo{volume}{120}},
  \bibinfo{pages}{067003} (\bibinfo{year}{2018}),
  \urlprefix\url{https://link.aps.org/doi/10.1103/PhysRevLett.120.067003}.

\bibitem[{\citenamefont{Altland and Zirnbauer}(1997)}]{Altland1997}
\bibinfo{author}{\bibfnamefont{A.}~\bibnamefont{Altland}} \bibnamefont{and}
  \bibinfo{author}{\bibfnamefont{M.~R.} \bibnamefont{Zirnbauer}},
  \bibinfo{journal}{Phys. Rev. B} \textbf{\bibinfo{volume}{55}},
  \bibinfo{pages}{1142} (\bibinfo{year}{1997}),
  \urlprefix\url{http://link.aps.org/doi/10.1103/PhysRevB.55.1142}.

\bibitem[{\citenamefont{Shastry and Shraiman}(1990)}]{Shastry1990}
\bibinfo{author}{\bibfnamefont{B.~S.} \bibnamefont{Shastry}} \bibnamefont{and}
  \bibinfo{author}{\bibfnamefont{B.~I.} \bibnamefont{Shraiman}},
  \bibinfo{journal}{Phys. Rev. Lett.} \textbf{\bibinfo{volume}{65}},
  \bibinfo{pages}{1068} (\bibinfo{year}{1990}),
  \urlprefix\url{https://link.aps.org/doi/10.1103/PhysRevLett.65.1068}.

\bibitem[{\citenamefont{Shastry and Shraiman}(1991)}]{Shastry1991}
\bibinfo{author}{\bibfnamefont{B.~S.} \bibnamefont{Shastry}} \bibnamefont{and}
  \bibinfo{author}{\bibfnamefont{B.~I.} \bibnamefont{Shraiman}},
  \bibinfo{journal}{Int. J. Mod. Phys. B} \textbf{\bibinfo{volume}{05}},
  \bibinfo{pages}{365} (\bibinfo{year}{1991}), ISSN \bibinfo{issn}{0217-9792},
  \urlprefix\url{https://doi.org/10.1142/S0217979291000237}.

\bibitem[{\citenamefont{Lake et~al.}(2022)\citenamefont{Lake, Patri, and
  Senthil}}]{Lake2022}
\bibinfo{author}{\bibfnamefont{E.}~\bibnamefont{Lake}},
  \bibinfo{author}{\bibfnamefont{A.~S.} \bibnamefont{Patri}}, \bibnamefont{and}
  \bibinfo{author}{\bibfnamefont{T.}~\bibnamefont{Senthil}},
  \bibinfo{journal}{Phys. Rev. B} \textbf{\bibinfo{volume}{106}},
  \bibinfo{pages}{104506} (\bibinfo{year}{2022}),
  \urlprefix\url{https://link.aps.org/doi/10.1103/PhysRevB.106.104506}.

\bibitem[{\citenamefont{Stewart}(2017)}]{Stewart2017}
\bibinfo{author}{\bibfnamefont{G.~R.} \bibnamefont{Stewart}},
  \bibinfo{journal}{Advances in Physics} \textbf{\bibinfo{volume}{66}},
  \bibinfo{pages}{75} (\bibinfo{year}{2017}), ISSN \bibinfo{issn}{0001-8732},
  \urlprefix\url{https://doi.org/10.1080/00018732.2017.1331615}.

\bibitem[{\citenamefont{Maeno et~al.}(2012)\citenamefont{Maeno, Kittaka,
  Nomura, Yonezawa, and Ishida}}]{Maeno2012}
\bibinfo{author}{\bibfnamefont{Y.}~\bibnamefont{Maeno}},
  \bibinfo{author}{\bibfnamefont{S.}~\bibnamefont{Kittaka}},
  \bibinfo{author}{\bibfnamefont{T.}~\bibnamefont{Nomura}},
  \bibinfo{author}{\bibfnamefont{S.}~\bibnamefont{Yonezawa}}, \bibnamefont{and}
  \bibinfo{author}{\bibfnamefont{K.}~\bibnamefont{Ishida}},
  \bibinfo{journal}{J. Phys. Soc. Jpn.} \textbf{\bibinfo{volume}{81}},
  \bibinfo{pages}{011009} (\bibinfo{year}{2012}), ISSN
  \bibinfo{issn}{0031-9015},
  \urlprefix\url{https://doi.org/10.1143/JPSJ.81.011009}.

\bibitem[{\citenamefont{Liu and Mao}(2015)}]{Liu2015a}
\bibinfo{author}{\bibfnamefont{Y.}~\bibnamefont{Liu}} \bibnamefont{and}
  \bibinfo{author}{\bibfnamefont{Z.-Q.} \bibnamefont{Mao}},
  \bibinfo{journal}{Physica C: Superconductivity and its Applications}
  \textbf{\bibinfo{volume}{514}}, \bibinfo{pages}{339} (\bibinfo{year}{2015}),
  ISSN \bibinfo{issn}{0921-4534},
  \urlprefix\url{https://www.sciencedirect.com/science/article/pii/S0921453415000660}.

\bibitem[{\citenamefont{Avers et~al.}(2020)\citenamefont{Avers, Gannon, Kuhn,
  Halperin, Sauls, DeBeer-Schmitt, Dewhurst, Gavilano, Nagy, Gasser
  et~al.}}]{Avers2020}
\bibinfo{author}{\bibfnamefont{K.~E.} \bibnamefont{Avers}},
  \bibinfo{author}{\bibfnamefont{W.~J.} \bibnamefont{Gannon}},
  \bibinfo{author}{\bibfnamefont{S.~J.} \bibnamefont{Kuhn}},
  \bibinfo{author}{\bibfnamefont{W.~P.} \bibnamefont{Halperin}},
  \bibinfo{author}{\bibfnamefont{J.~A.} \bibnamefont{Sauls}},
  \bibinfo{author}{\bibfnamefont{L.}~\bibnamefont{DeBeer-Schmitt}},
  \bibinfo{author}{\bibfnamefont{C.~D.} \bibnamefont{Dewhurst}},
  \bibinfo{author}{\bibfnamefont{J.}~\bibnamefont{Gavilano}},
  \bibinfo{author}{\bibfnamefont{G.}~\bibnamefont{Nagy}},
  \bibinfo{author}{\bibfnamefont{U.}~\bibnamefont{Gasser}},
  \bibnamefont{et~al.}, \bibinfo{journal}{Nature Physics}
  \textbf{\bibinfo{volume}{16}}, \bibinfo{pages}{531} (\bibinfo{year}{2020}),
  ISSN \bibinfo{issn}{1745-2481},
  \urlprefix\url{https://doi.org/10.1038/s41567-020-0822-z}.

\bibitem[{\citenamefont{Aoki et~al.}(2022)\citenamefont{Aoki, Brison, Flouquet,
  Ishida, Knebel, Tokunaga, and Yanase}}]{Aoki2022}
\bibinfo{author}{\bibfnamefont{D.}~\bibnamefont{Aoki}},
  \bibinfo{author}{\bibfnamefont{J.-P.} \bibnamefont{Brison}},
  \bibinfo{author}{\bibfnamefont{J.}~\bibnamefont{Flouquet}},
  \bibinfo{author}{\bibfnamefont{K.}~\bibnamefont{Ishida}},
  \bibinfo{author}{\bibfnamefont{G.}~\bibnamefont{Knebel}},
  \bibinfo{author}{\bibfnamefont{Y.}~\bibnamefont{Tokunaga}}, \bibnamefont{and}
  \bibinfo{author}{\bibfnamefont{Y.}~\bibnamefont{Yanase}},
  \bibinfo{journal}{Journal of Physics: Condensed Matter}
  \textbf{\bibinfo{volume}{34}}, \bibinfo{pages}{243002}
  (\bibinfo{year}{2022}), ISSN \bibinfo{issn}{0953-8984},
  \urlprefix\url{https://dx.doi.org/10.1088/1361-648X/ac5863}.

\bibitem[{\citenamefont{Mattis and Bardeen}(1958)}]{Mattis1958}
\bibinfo{author}{\bibfnamefont{D.~C.} \bibnamefont{Mattis}} \bibnamefont{and}
  \bibinfo{author}{\bibfnamefont{J.}~\bibnamefont{Bardeen}},
  \bibinfo{journal}{Phys. Rev.} \textbf{\bibinfo{volume}{111}},
  \bibinfo{pages}{412} (\bibinfo{year}{1958}),
  \urlprefix\url{https://link.aps.org/doi/10.1103/PhysRev.111.412}.

\bibitem[{\citenamefont{Leplae}(1983)}]{Leplae1983}
\bibinfo{author}{\bibfnamefont{L.}~\bibnamefont{Leplae}},
  \bibinfo{journal}{Phys. Rev. B} \textbf{\bibinfo{volume}{27}},
  \bibinfo{pages}{1911} (\bibinfo{year}{1983}),
  \urlprefix\url{https://link.aps.org/doi/10.1103/PhysRevB.27.1911}.

\bibitem[{\citenamefont{Zimmermann et~al.}(1991)\citenamefont{Zimmermann,
  Brandt, Bauer, Seider, and Genzel}}]{Zimmermann1991}
\bibinfo{author}{\bibfnamefont{W.}~\bibnamefont{Zimmermann}},
  \bibinfo{author}{\bibfnamefont{E.}~\bibnamefont{Brandt}},
  \bibinfo{author}{\bibfnamefont{M.}~\bibnamefont{Bauer}},
  \bibinfo{author}{\bibfnamefont{E.}~\bibnamefont{Seider}}, \bibnamefont{and}
  \bibinfo{author}{\bibfnamefont{L.}~\bibnamefont{Genzel}},
  \bibinfo{journal}{Physica C: Superconductivity}
  \textbf{\bibinfo{volume}{183}}, \bibinfo{pages}{99} (\bibinfo{year}{1991}),
  ISSN \bibinfo{issn}{0921-4534},
  \urlprefix\url{https://www.sciencedirect.com/science/article/pii/092145349190771P}.

\bibitem[{\citenamefont{Bradley and Davies}(1968)}]{Bradley1968}
\bibinfo{author}{\bibfnamefont{C.}~\bibnamefont{Bradley}} \bibnamefont{and}
  \bibinfo{author}{\bibfnamefont{B.}~\bibnamefont{Davies}},
  \bibinfo{journal}{Rev. Mod. Phys.} \textbf{\bibinfo{volume}{40}},
  \bibinfo{pages}{359} (\bibinfo{year}{1968}),
  \urlprefix\url{https://link.aps.org/doi/10.1103/RevModPhys.40.359}.

\end{thebibliography}

\end{document}